\documentclass[12pt]{article}
\usepackage{epsfig}
\date{}
\begin{document}

{\large \sf
\title{
{\normalsize
\begin{flushright}
\end{flushright}}
{\vspace{2cm} {\LARGE \sf A Timeon Model of Quark and Lepton Mass
Matrices\thanks{This research was supported in part by the U.S.
Department of Energy (grant no. DE-FG02-92-ER40699)}} \vspace{2cm}}
}

{\large \sf
\author{
{\large \sf
R. Friedberg$^1$ and  T. D. Lee$^{1,~2}$}\\
{\normalsize \it 1. Physics Department, Columbia University}\\
{\normalsize \it New York, NY 10027, U.S.A.}\\
{\normalsize \it 2. China Center of Advanced Science and Technology (CCAST/World Lab.)}\\
{\normalsize \it P.O. Box 8730, Beijing 100190, China}\\
} \maketitle

\newpage

\begin{abstract}

{\normalsize \sf

It is proposed that $T$ violation in physics, as well as the masses
of electron and $u,~d$ quarks, arise from a pseudoscalar interaction
with a new spin $0$ field $\tau(x)$, odd in $P$ and $T$, but even in
$C$. This interaction contains a factor $i\gamma_5$ in the quark and
lepton Dirac algebra, so that the full Hamiltonian is $P$, $T$
conserving; but by spontaneous symmetry breaking, the new field
$\tau(x)$ has a nonzero expectation value $<\tau>\neq 0$ that breaks
$P$ and $T$ symmetry. Oscillations of $\tau(x)$ about its
expectation value produce a new particle, the "timeon". The mass of
timeon is expected to be high because of its flavor-changing
properties.

The main body of the paper is on the low energy phenomenology of
the timeon model. As we shall show, for the quark system the model
gives a compact $3$-dimensional geometric picture consisting of
two elliptic plates and one needle, which embodies the ten
observables: six quark masses, three Eulerian angles
$\theta_{12},~\theta_{23},~\theta_{31}$ and the Jarlskog invariant
of the CKM matrix.

For leptons, we assume that the neutrinos do not have a direct
timeon interaction; therefore, the lowest neutrino mass is zero. The
timeon interaction with charged leptons yields the observed nonzero
electron mass, analogous to the up and down quark masses.
Furthermore, the timeon model for leptons contains two fewer
theoretical parameters than observables. Thus, there are two
testable relations between the three angles
$\theta_{12},~\theta_{23},~\theta_{31}$ and the Jarlskog invariant
of the neutrino mapping matrix.}
\end{abstract}

{\normalsize \sf PACS{:~~12.15.Ff,~~11.30.Er}}

\vspace{1cm}


\newpage

\section*{\Large \sf  1. Introduction}
\setcounter{section}{1} \setcounter{equation}{0}

We suggest that the observed $CP$ and $T$ violations are due to a
new $P$ odd and $T$ odd spin zero field $\tau(x)$, called the timeon
field; the same field is also responsible for the small masses of
$u,~d$ quarks, as well as that of the electron. Consider first the
quark system. Let $q_i(\uparrow)$ and $q_i(\downarrow)$ be the quark
states "diagonal" in $W^\pm$ transitions[1]:
$$
q_i(\downarrow)\rightleftharpoons q_i(\uparrow) + W^-\eqno(1.1)
$$
and
$$
q_i(\uparrow)\rightleftharpoons q_i(\downarrow) + W^+\eqno(1.2)
$$
with $i=1,~2$ and $3$. The electric charges in units of $e$ are
$+\frac{2}{3}$ for $q_i(\uparrow)$ and $-\frac{1}{3}$ for
$q_i(\downarrow)$. These quark states $q_i(\uparrow)$ and
$q_i(\downarrow)$ are, however, not the mass eigenstates $d,~s,~b$
and $u,~c,~t$. We assume that the mass Hamiltonians $H_\uparrow$ for
$q_i(\uparrow)$ and $H_\downarrow$ for $q_i(\downarrow)$ are given
by
$$
H_{\uparrow/\downarrow}=\bigg(q_1^\dag,~
q_2^\dag,~q_3^\dag\bigg)_{\uparrow/\downarrow} (G \gamma_4 +
iF\gamma_4 \gamma_5)_{\uparrow/\downarrow}\left(
\begin{array}{r}
q_1\\
q_2\\
q_3
\end{array}\right)_{\uparrow/\downarrow}\eqno(1.3)
$$
with the $3\times 3$ matrices $G_{\uparrow/\downarrow}$ and
$F_{\uparrow/\downarrow}$ both real and hermitian. The mass matrix
$G_{\uparrow/\downarrow}$ is the same zeroth order mass matrix as
${\cal M}_0(q_{\uparrow/\downarrow})$ of Ref.~1, given by
$$
G_{\uparrow/\downarrow}=\left(
\begin{array}{ccc}
\beta \eta^2(1+\xi^2)& -\beta \eta & -\beta \xi \eta\\
-\beta \eta &\beta + \alpha \xi^2 & -\alpha \xi\\
-\beta \xi \eta &-\alpha \xi& \alpha +\beta
\end{array}\right)_{\uparrow/\downarrow}, \eqno(1.4)
$$
in which $\alpha_\uparrow,~\beta_\uparrow,~
\xi_\uparrow,~\eta_\uparrow$ and
$\alpha_\downarrow,~\beta_\downarrow,~
\xi_\downarrow,~\eta_\downarrow$ are all real parameters with
$\alpha_{\uparrow/\downarrow}$ and $\beta_{\uparrow/\downarrow}$ to
be positive. It can be readily verified that the determinants
$$
|G_\uparrow|=|G_\downarrow|=0.\eqno(1.5)
$$
Thus, the lowest eigenvalues of $G_\uparrow$ and $G_\downarrow$ are
both zero. These two real symmetric matrices can be diagonalized by
real, orthogonal matrices $(U_\uparrow)_0$ and $(U_\downarrow)_0$,
with
$$
(U_\uparrow)_0^\dag G_\uparrow (U_\uparrow)_0=\left(
\begin{array}{ccc}
0 & 0 & 0\\
0 & m_0(c) & 0\\
0 & 0 & m_0(t)
\end{array}\right) \eqno(1.6)
$$
and
$$
(U_\downarrow)_0^\dag G_\downarrow (U_\downarrow)_0=\left(
\begin{array}{ccc}
0 & 0 & 0\\
0 & m_0(s) & 0\\
0 & 0 & m_0(b)
\end{array}\right) \eqno(1.7)
$$
where the nonzero eigenvalues are the zeroth order masses of $c,~t$
and $s,~b$ quarks, with
$$
m_0(c)=\beta_\uparrow[1+\eta_\uparrow^2(1+\xi_\uparrow^2)],\eqno(1.8)
$$
$$
m_0(t)= \alpha_\uparrow(1+\xi_\uparrow^2)+\beta_\uparrow,\eqno(1.9)
$$
$$
m_0(s)=\beta_\downarrow[1+\eta_\downarrow^2(1+\xi_\downarrow^2)]\eqno(1.10)
$$
and
$$
 m_0(b)=\alpha_\downarrow(1+\xi_\downarrow^2)+\beta_\downarrow.\eqno(1.11)
$$
Thus, $G_\uparrow$ and $G_\downarrow$ can be each represented by an
ellipse of minor and major axes given by $m_0(c)$ and $m_0(t)$ for
$\uparrow$ and likewise $m_0(s)$ and $m_0(b)$ for $\downarrow$.

The orientations of these two elliptic plates are determined by
their eigenstates. As in Ref.~1, we define four real angular
variables $\theta_\downarrow,~\phi_\downarrow$ and
$\theta_\uparrow,~\phi_\uparrow$ by
$$
\xi_\downarrow=\tan \phi_\downarrow,~~\xi_\uparrow=\tan
\phi_\uparrow
$$
$$
\eta_\downarrow=\tan \theta_\downarrow \cos \phi_\downarrow~~{\sf
and}~~\eta_\uparrow=\tan \theta_\uparrow \cos \phi_\uparrow.
\eqno(1.12)
$$
The eigenstates of $G_\uparrow$ are
$$
\epsilon_\uparrow = \left(
\begin{array}{l}
\cos \theta_\uparrow\\
\sin \theta_\uparrow \cos \phi_\uparrow\\
\sin \theta_\uparrow \sin \phi_\uparrow
\end{array}
\right)~{\sf with~eigenvalue}~0, \eqno(1.13)
$$
$$
p_\uparrow = \left(
\begin{array}{l}
-\sin \theta_\uparrow\\
\cos \theta_\uparrow \cos \phi_\uparrow\\
\cos \theta_\uparrow \sin \phi_\uparrow
\end{array}
\right)~{\sf with~eigenvalue}~m_0(c) \eqno(1.14)
$$
and
$$
P_\uparrow = \left(
\begin{array}{l}
~~~~0\\
-\sin \phi_\uparrow\\
~~\cos \phi_\uparrow
\end{array}
\right)~{\sf with~eigenvalue}~m_0(t). \eqno(1.15)
$$
Correspondingly, the eigenstates of $G_\downarrow$ are
$$
\epsilon_\downarrow = \left(
\begin{array}{l}
~~\cos \theta_\downarrow\\
-\sin \theta_\downarrow \cos \phi_\downarrow\\
-\sin \theta_\downarrow \sin \phi_\downarrow
\end{array}
\right)~{\sf with~eigenvalue}~0, \eqno(1.16)
$$
$$
p_\downarrow = \left(
\begin{array}{l}
\sin \theta_\downarrow\\
\cos \theta_\downarrow \cos \phi_\downarrow\\
\cos \theta_\downarrow \sin \phi_\downarrow
\end{array}
\right)~{\sf with~eigenvalue}~m_0(s), \eqno(1.17)
$$
and
$$
P_\downarrow = \left(
\begin{array}{l}
~~~~0\\
-\sin \phi_\downarrow\\
~~\cos \phi_\downarrow
\end{array}
\right)~{\sf with~eigenvalue}~m_0(b). \eqno(1.18)
$$
We note that by changing $\theta_\uparrow$, $\phi_\uparrow$ to
$-\theta_\downarrow$, $\phi_\downarrow$ the unit vectors
$\epsilon_\uparrow$, $p_\uparrow$, $P_\uparrow$ of (1.13)-(1.15)
become $\epsilon_\downarrow$, $p_\downarrow$ and $P_\downarrow$ of
(1.16)-(1.18). Here the signs of $\theta_\uparrow$ and
$\theta_\downarrow$ are chosen so that the sign convention of the
particle data group's CKM matrix agrees with both $\theta_\uparrow$
and $\theta_\downarrow$ being positive, as we shall see. In terms of
these eigenstates, the $3\times 3$ real unitary matrices
$(U_\uparrow)_0$ and $(U_\downarrow)_0$ of (1.6)-(1.7) are given by
$$
(U_\uparrow)_0=(\epsilon_\uparrow,~p_\uparrow,~P_\uparrow).\eqno(1.19)
$$
and
$$
(U_\downarrow)_0=(\epsilon_\downarrow,~p_\downarrow,~P_\downarrow)\eqno(1.20)
$$
Thus, in the absence of the $iF_{\uparrow/\downarrow}
\gamma_4\gamma_5$ term in (1.3), the corresponding CKM matrix in
this approximation is given by
$$
(U_{CKM})_0=(U_\uparrow)_0^\dag (U_\downarrow)_0=
$$
$$\left(
\begin{array}{ccc}
\cos \theta_\downarrow\cos \theta_\uparrow &\sin\theta_\downarrow
\cos\theta_\uparrow &\sin\theta_\uparrow \sin\phi\\
~~-\sin\theta_\downarrow \sin\theta_\uparrow \cos\phi
&~~+\cos\theta_\downarrow \sin\theta_\uparrow \cos\phi &\\
&&\\
-\cos \theta_\downarrow\sin \theta_\uparrow &-\sin\theta_\downarrow
\sin\theta_\uparrow &\cos\theta_\uparrow \sin\phi\\
~~-\sin\theta_\downarrow \cos\theta_\uparrow \cos\phi
&~~+\cos\theta_\downarrow \cos\theta_\uparrow \cos\phi &\\
&&\\
 \sin\theta_\downarrow \sin\phi
&-\cos\theta_\downarrow\sin\phi&\cos\phi
\end{array}
 \right )~,\eqno(1.21)
$$
in which
$$
\phi=\phi_\uparrow-\phi_\downarrow.\eqno(1.22)
$$

We assume that $T$ violation and the small masses of $u,~d$ quarks
are due to the new
$$
iF\gamma_4\gamma_5\eqno(1.23)
$$
term in (1.3), with
$$
F_\uparrow=F_\downarrow=F=\tau_q f\tilde{f}\eqno(1.24)
$$
in which $\tau_q$ is a real constant and $f$ a $3$ dimensional unit
vector represented by its $3\times 1$ real column matrix.
Graphically, we can visualize $G_\uparrow$ and $G_\downarrow$ as two
elliptic plates mentioned above, and $F_{\uparrow/\downarrow}$ as a
single needle of length $\tau_q$ and direction $f$, as shown in
Figure~1.

We note that the time-reversal operation $T$ in quantum mechanics
involves a complex conjugation operation changing the factor $i$ to
$-i$. Since the entirety of classical mechanics can be formulated
with real numbers, the presence of $i$ in quantum mechanics is
necessitated by the commutation or anticommutation relation between
operators, such as that between $\gamma_4$ and $\gamma_5$. This led
us to postulate (1.23) as the source of $T$ violation. The specific
form given by (1.23)-(1.24) can be due to the spontaneous symmetry
breaking of a new $T$ odd, $P$ odd and $CP$ odd, spin $0$ field
$\tau(x)$, which has a vacuum expectation value given by
$$
<\tau(x)>_{vac}=\tau_q\neq 0.\eqno(1.25)
$$
While the general characteristics of spontaneous time reversal
symmetry breaking models have  been discussed in the literature[2],
one of the new features of the present model is to connect such
symmetry breaking with the smallness of the light quark and electron
masses.

In Section 2, we begin with a general $3\times 3$ $T$, $P$ and $CP$
odd mass matrix of the form
$$
{\cal G} \gamma_4 + i{\cal F}\gamma_4 \gamma_5\eqno(1.26)
$$
with ${\cal G}$ and ${\cal F}$ both real and hermitian, then derive
some useful properties of its eigenvalues and eigenvectors. In
Sections 3 and 4, we summarize the analysis of how in (1.24), the
length $\tau_q$ and the direction $f$ of the needle are related to
the light quark masses and the Jarlskog invariant[3] ${\cal J}$ of
the CKM matrix[4,5]. As we shall see, this leads to
$$
\tau_q\cong 33 MeV,\eqno(1.27)
$$
$$
m_u\cong \tau_q(\tilde{f}\epsilon_\uparrow)^2\eqno(1.28)
$$
and
$$
m_d\cong \tau_q(\tilde{f}\epsilon_\downarrow)^2\eqno(1.29)
$$
with $\epsilon_\uparrow$ and $\epsilon_\downarrow$ given by (1.13)
and (1.16).

An interesting feature of the model is: the implicit assumption that
the constant $\tau_q$ might be due to the spontaneous symmetry
breaking of a new type of $T$ odd and $CP$ odd, spin $0$ field
$\tau(x)$, which has a vacuum expectation value given by (1.25). As
an example, we may assume that the Lagrangian density of $\tau(x)$
is given by
$$
-\frac{1}{2}\bigg(\frac{\partial \tau}{\partial
x_\mu}\bigg)^2-V(\tau)\eqno(1.30)
$$
with
$$
V(\tau)=-\frac{1}{2}\lambda
\tau^2(\tau_q^2-\frac{1}{2}\tau^2)\eqno(1.31)
$$
in which the (renormalized) value of $\lambda$ is positive. This
then yields (1.25). Expanding $V(\tau)$ around its equilibrium value
$\tau=\tau_q$, we have
$$
V(\tau)=-\frac{\lambda}{4}\tau_q^4+\frac{1}{2}m_\tau^2(\tau-\tau_q)^2
+O[(\tau-\tau_q)^3]\eqno(1.32)
$$
with
$$
m_\tau=(2\lambda)^{\frac{1}{2}}\tau_q,\eqno(1.33)
$$
the mass of this new $T$ violating, $C$ violating and $CP$ violating
quantum, called timeon.

The interaction between $\tau(x)$ and the quark field might be
obtained by replacing the $F=\tau_q f\tilde{f}$ factor of (1.24)
with
$$
F=\tau(x)f\tilde{f}.\eqno(1.34)
$$
Because of the flavor-changing property of the timeon field[6], its
mass (if it exists) must be quite high. A full analysis of this
interesting possibility lies outside the scope of this paper. Here,
we concentrate on the low energy phenomenology of the timeon model.

In the application to quarks, there are ten measurable parameters in
$H_{\uparrow/\downarrow}$ given by (1.3). These consist of 3 angles
$$
\theta_\uparrow,~\theta_\downarrow~{\sf
and}~\phi=\phi_\uparrow-\phi_\downarrow\eqno(1.35)
$$
of (1.21) and 4 zeroth order masses
$$
m_0(c),~m_0(t),~m_0(s)~{\sf and}~m_0(b)\eqno(1.36)
$$
given by (1.8)-(1.11) in the description of $G_\uparrow$ and
$G_\downarrow$. In addition, the timeon term (1.24) contains 3
parameters:
$$
\tau_q~{\sf and~two~angles~in}~f.\eqno(1.37)
$$
(The angle $\phi_\uparrow+\phi_\downarrow$ is an unphysical gauge
parameter.) These ten theoretical parameters account for ten
observables: six quark masses, three Eulerian angles
$$
\theta_{12},~\theta_{23},~\theta_{31}\eqno(1.38)
$$
and the $T$-violating phase factor
$$
e^{i\delta}\eqno(1.39)
$$
in the CKM matrix. The timeon model provides a simple picture,
combining these ten observables into a single compact geometric
structure.

In Section~5, we extend the timeon model to leptons. Similar to
quarks, there are also 10 observables: six leptonic masses and
four angles (as in (1.38) and (1.39)) of the neutrino mapping
matrix. However, unlike the $\uparrow$ and $\downarrow$ quarks,
the mass scales of the neutrinos are much smaller than those of
the charged leptons. Thus, it seems reasonable to explore the
interesting possibility that the timeon term is \underline{absent}
in the neutrino sector. In this sense, the neutrino sector may be
regarded as more "primeval", with its lowest neutrino mass zero.
As we shall discuss, the corresponding timeon model for leptons
gives electron a mass and the number of parameters in the theory
can be reduced to only 8. Thus, there are 2 testable relations
between the 10 observables.

\newpage

\section*{\Large \sf 2. Eigenvalues and Eigenvectors}

\noindent{\bf 2.1 General Formulation}\\

We begin with (1.26) and write the corresponding mass Hamiltonian as
$$
H=\psi^\dag ({\cal G} \gamma_4 + i {\cal F} \gamma_4\gamma_5)
\psi\eqno(2.1)
$$
where ${\cal G}$ and ${\cal F}$ are both $3\times 3$ real hermitian
matrices. Resolve the Dirac field operator $\psi$ into a sum of
left-handed and right-handed components:
$$
\psi=L+R\eqno(2.2)
$$
with
$$
L=\frac{1}{2}(1+\gamma_5)\psi~~~{\sf
and}~~~R=\frac{1}{2}(1-\gamma_5)\psi.\eqno(2.3)
$$
Thus, (2.1) becomes
$$
H=L^\dag ({\cal G}-i{\cal F}) \gamma_4 R + R^\dag({\cal G} + i {\cal
F}) \gamma_4 L.\eqno(2.4)
$$
Define
$$
M\equiv {\cal G} -i{\cal F},\eqno(2.5)
$$
$$
M^\dag \equiv {\cal G} +i{\cal F}\eqno(2.6)
$$
and their product
$$
{\cal M}^2\equiv MM^\dag=({\cal G} -i{\cal F})({\cal G} +i{\cal
F}).\eqno(2.7)
$$
Since ${\cal M}^2$ is hermitian, it can be diagonalized by a unitary
transformation. We write
$$
V_L^\dag MM^\dag V_L=m^2_D={\sf diagonal}\eqno(2.8)
$$
with
$$
V_L^\dag V_L=1.
$$
Multiply (2.8) on the right by $m_D^{-1}$, we see that by defining
$$
V_R\equiv M^\dag V_Lm_D^{-1},\eqno(2.9)
$$
we have
$$
V_L^\dag M V_R=m_D;\eqno(2.10)
$$
furthermore, $V_R$ is also the unitary matrix that can diagonalize
the corresponding $M^\dag M$; i.e.,
$$
V_R^\dag M^\dag M V_R=m^2_D.\eqno(2.11)
$$
(Note that ${\cal M}^2=M M^\dag$, but ${\cal M}$ can be quite
different from $M$ or $M^\dag$.)

For application, the matrices ${\cal G}$ and ${\cal F}$ can be
either $G_\uparrow$, $F_\uparrow$ or $G_\downarrow$, $F_\downarrow$
of (1.3). As in (1.13)-(1.15), we write the eigenstates of ${\cal
G}$ as column matrices
$$
\epsilon,~p~{\sf and}~P.\eqno(2.12)
$$
Likewise, write the unit column matrix $f$ in (1.24) as
$$
f= \left(
\begin{array}{l}
\cos a\\
\sin a \cos b\\
\sin a \sin b
\end{array}
\right).\eqno(2.13)
$$
Thus,
$$
{\cal G}=\nu\epsilon\tilde{\epsilon}+\mu p\tilde{p} +m
P\tilde{P}\eqno(2.14)
$$
and correspondingly we may set
$$
{\cal F}=\tau f \tilde{f},\eqno(2.15)
$$
with $\nu,~\mu,~m,~\tau$ all real constants. When $\tau=\tau_q$
and $\nu=0$, ${\cal F}$ becomes $F$ of (1.24) and ${\cal G}$ can
be either $G_\uparrow$ or $G_\downarrow$ of (1.4). For the moment,
we retain the eigenvalue $\nu$ in (2.14) for the formal symmetry
of some of the mathematical expressions (as in (2.23) below), even
though $\nu=0$ when we discuss physical applications of our model.

As in (1.19) and (1.20), we define a real unitary matrix $U_0$ whose
columns are $\epsilon,~p$ and $P$ of (2.12); i.e.,
$$
U_0= (\epsilon~~p~~P).\eqno(2.16)
$$
The matrix $U_0$ diagonalizes ${\cal G}$, with
$$
{\cal G}' \equiv\tilde{U}_0{\cal G}U_0= \left(
\begin{array}{ccc}
\nu & 0 & 0\\
0 &\mu & 0\\
0 & 0 & m
\end{array}
\right).\eqno(2.17)
$$
It also transforms ${\cal F}$ into
$$
{\cal F}'\equiv \tilde{U}_0{\cal F}U_0 = \tau
f'\tilde{f}'\eqno(2.18)
$$
where
$$
f'=\left(\begin{array}{l}
f_\epsilon\\
f_p\\
f_P
\end{array}
\right)\eqno(2.19)
$$
with
$$
f_\epsilon
=\tilde{\epsilon}f,~~f_p=\tilde{p}f,~~~f_P=\tilde{P}f.\eqno(2.20)
$$
As before,
$$
f_\epsilon^2+f_p^2+f_P^2=1.\eqno(2.21)
$$

By using (2.17)-(2.19), we find that the same $U_0$ also transforms
the matrix ${\cal M}^2$ into
$$
({\cal M}')^2 = \tilde{U}_0{\cal M}^2U_0=({\cal G}')^2+({\cal
F}')^2+i[{\cal G}'~,{\cal F}'],\eqno(2.22)
$$
which is given by
$$
({\cal M}')^2  = \left(
\begin{array}{ccc}
\nu^2+\tau^2 f_\epsilon^2 & \tau[\tau-i(\mu-\nu)]f_\epsilon f_p &
\tau[\tau-i(m-\nu)]f_Pf_\epsilon\\
\tau[\tau+i(\mu-\nu)]f_\epsilon f_p&\mu^2+\tau^2f_p^2 &
\tau[\tau-i(m-\mu)]f_pf_P\\
\tau[\tau+i(m-\nu)]f_Pf_\epsilon & \tau[\tau+i(m-\mu)]f_pf_P &
m^2+\tau^2f_P^2
\end{array}
\right). \eqno(2.23)
$$
For our applications, we are only interested in the case $\nu=0$.
Define
$$
{\cal N} \equiv \lim_{\nu=0}({\cal M}')^2\eqno(2.24)
$$
and let $\lambda_1^2,~\lambda_2^2,~\lambda_3^2$ be the eigenvalues
of ${\cal N}$. From (2.23) and (2.24) we have
$$
\lambda_1^2+\lambda_2^2+\lambda_3^2=m^2+\mu^2+\tau^2\eqno(2.25)
$$
and
$$
\lambda_1^2\lambda_2^2\lambda_3^2=|{\cal
N}|=\tau^2f_\epsilon^4\mu^2m^2,\eqno(2.26)
$$
These eigenvalues are also the solution $\lambda^2$ of the cubic
equation
$$
|{\cal N}-\lambda^2|=|{\cal
N}|+A\lambda^2+B\lambda^4-\lambda^6=0\eqno(2.27)
$$
where
$$
A=-\mu^2m^2-\tau^2[m^2(1-f_P^2)^2+\mu^2(1-f_p^2)^2+2m\mu
f_p^2f_P^2]\eqno(2.28)
$$
and
$$
B=m^2+\mu^2+\tau^2.\eqno(2.29)
$$
In the limit $\tau\rightarrow 0$, we see from (2.25)-(2.29) that the
two heavier masses become $\mu$ and $m$, while the lightest mass is
proportional to $\tau$. (Readers who are only interested in
perturbative solutions are encouraged to move on to Section 3
directly.)\\

\noindent{\bf 2.2 Some Useful Expressions}\\

It is convenient to arrange the three eigenvalues
$\lambda_1^2,~\lambda_2^2,~\lambda_3^2$ in an ascending order, each
with a new subscript:
$$
\lambda_s^2<\lambda_l^2<\lambda_L^2,\eqno(2.30)
$$
Write
$$
E_i=\lambda_i^2\eqno(2.31)
$$
with
$$
i=s,~l~~{\sf and}~~L.\eqno(2.32)
$$
(Here, the letters are $s$ for small, $l$ for large and $L$ for
very large.) Let $\psi_i$ be the corresponding eigenstate defined
by
$$
{\cal N}\psi_i=E_i\psi_i.\eqno(2.33)
$$
Using (2.23)-(2.24), we can express ${\cal N}$ in the form
$$
{\cal N}=\left(
\begin{array}{cc}
h & n\chi\\
n\chi^\dag & n^2
\end{array}\right) \eqno(2.34)
$$
where
$$
h=\left(
\begin{array}{cc}
\tau^2f_\epsilon^2 & \tau(\tau-i\mu)f_\epsilon f_p\\
\tau(\tau+i\mu)f_\epsilon f_p& \mu^2+\tau^2f_p^2
\end{array}\right), \eqno(2.35)
$$
$$
n\chi=\left(
\begin{array}{c}
(\tau-im)f_\epsilon \\
(\tau-i(m-\mu)) f_p
\end{array}\right)\tau f_P \eqno(2.36)
$$
and
$$
n^2=m^2+\tau^2f_P^2,\eqno(2.37)
$$
Correspondingly, each eigenvector $\psi_i$ of ${\cal N}$ can be
written as
$$
\psi_i=\left(
\begin{array}{c}
\phi_i\\
c_i
\end{array}\right)\eqno(2.38)
$$
with $\phi_i$ a $2\times 1$ column matrix and $c_i$ a constant. From
(2.33), (2.34) and (2.38), we have
$$
h\phi_i+(nc_i)\chi=E_i\phi_i\eqno(2.39)
$$
and
$$
n\chi^\dag\phi_i+n^2c_i=E_ic_i.\eqno(2.40)
$$
From (2.39), it follows that
$$
\phi_i=(E_i-h)^{-1}n c_i\chi,\eqno(2.41)
$$
and likewise from (2.40),
$$
c_i=(E_i-n^2)^{-1}n\chi^\dag\phi_i.\eqno(2.42)
$$
Substituting (2.41) to (2.40) we find
$$
m_i^2\equiv E_i=n^2[1+\chi^\dag(m_i^2-h)^{-1}\chi].\eqno(2.43)
$$
Likewise, (2.39) and (2.42) lead to
$$
(h-\frac{n^2}{n^2-E_i}\chi\chi^\dag)\phi_i=E_i\phi_i.\eqno(2.44)
$$
Both (2.43) and (2.44) are valid for all three solutions $i=s,~l$
and $L$. From (2.44), we see that once the eigenvalue $E_i$ is
known, the determination of the corresponding three-dimensional
eigenvector $\psi_i$ reduces to the much simpler two-dimensional
spinor equation (2.44). This fact will be of use in the later
sections on leptons.\\

\noindent{\bf \underline{Remarks}} It is well known that a mass
matrix of the form (2.4) can also be written as a single term
$$
\psi^\dag M\gamma_4\psi,\eqno(2.45)
$$
but with the hermitian matrix $M$ complex. However, the reality
conditions of ${\cal G}$ and ${\cal F}$ lead to a specific form of
$M$, and that specific form may appear "unnatural" without the
"timeon" picture, Further discussions will be given in Appendix A.\\

\noindent{\bf 2.3 Determination of Eigenvalues}\\

For quarks, the top mass $m_t$ is much larger than $m_c$ and $m_u$,
so is the bottom mass $m_b>>m_s$ and $m_d$. Likewise for charged
leptons, $m_\tau$ is $>>m_\mu$ and $m_e$. Thus, in the notations of
(2.32), for $i=L$ the largest mass $m_L$ satisfies
$$
m_L>>m_l>m_s.\eqno(2.46)
$$
Eq.(2.43) gives a convenient route to express $m_L$ in terms of the
parameters in ${\cal N}$, given by (2.34)-(2.37). In this case, we
can regard the parameter $m$ in (2.23) as satisfying
$$
m=O(m_L)>>\mu~~{\sf and}~~\tau.\eqno(2.47)
$$
Define the parameter $\epsilon$ through
$$
m_L^2=n^2+\epsilon=m^2+\tau^2f_P^2+\epsilon.\eqno(2.48)
$$
Eq.(2.43) in the case $i=L$ gives
$$
m_L^2=n^2+\chi^\dag\frac{n^2}{n^2+\epsilon-h}\chi\eqno(2.49)
$$
which leads to
$$
\epsilon=\chi^\dag\frac{n^2}{n^2+\epsilon-h}\chi=
\chi^\dag\bigg(1-\frac{h-\epsilon}{n^2}\bigg)^{-1}\chi
$$
$$
=\chi^\dag\chi+\chi^\dag\frac{h-\epsilon}{n^2}\chi+
\chi^\dag\bigg(\frac{h-\epsilon}{n^2}\bigg)^2\chi+\cdots.
$$
Thus, we have the expansion
$$
m_L^2=m^2+\tau^2f_P^2+\chi^\dag\chi+\frac{1}{m^2}[\chi^\dag
h\chi-(\chi^\dag\chi)^2]+O(m^{-4});\eqno(2.50)
$$
i.e., on account of (2.35)-(2.37)
$$
m_L^2=m^2+\tau^2f_P^2\bigg[1+f_\epsilon^2+\bigg(1-\frac{\mu}{m}\bigg)^2f_p^2\bigg]
+O\bigg(\frac{\tau^2\mu^2}{m^2}\bigg).\eqno(2.51)
$$

To derive the corresponding expansions for the smaller masses
$m_s^2$ and $m_l^2$, define
$$
\begin{array}{ll}
&{\cal S}\equiv m_s^2+m_l^2\\
 {\sf
and}~~~~~~~~~~~~~~~~~~~~~&~~~~~~~~~~~~~~~~~~~~~~~~~~~~~~~~~~
~~~~~~~~~~~~~~~~~~~~~~~~~~~(2.52)\\
&{\cal P}\equiv m_s^2m_l^2.
\end{array}
$$
Thus,
$$
\begin{array}{ll}
&m_l^2=\frac{1}{2}[{\cal S}+({\cal S}^2-4{\cal P})^{\frac{1}{2}}]~~~~~~~~~~~~~~~~~~~~~~~~~~~~~~\\
{\sf and}~~~~~~~~~~~~~~~~~~~~~&~~~~~~~~~~~~~~~~~~~~~~~~~~~~~~~~~~
~~~~~~~~~~~~~~~~~~~~~~~~~~~(2.53)\\
&m_s^2=\frac{1}{2}[{\cal S}-({\cal S}^2-4{\cal P})^{\frac{1}{2}}].
\end{array}
$$
From (2.25)-(2.26), we have
$$
\begin{array}{ll}
&{\cal S}=m^2+\mu^2+\tau^2-m_L^2~~~~~~~~~~~~~~~~~~~~~~~~~~~~~~~~~\\
{\sf and}~~~~~~~~~~~~~~~~~~~~~&~~~~~~~~~~~~~~~~~~~~~~~~~~~~~~~~~~
~~~~~~~~~~~~~~~~~~~~~~~~~~~(2.54)\\
&{\cal P}=\tau^2f_\epsilon^4\mu^2m^2/m_L^2.
\end{array}
$$
Combining these with (2.51), we have the expressions for $m_l^2$ and
$m_s^2$. In the limit $m\rightarrow \infty$, but $\tau$ comparable
to $\mu$, (2.54) becomes
$$
\begin{array}{ll}
&{\cal S}\rightarrow \mu^2+\tau^2(f_\epsilon^2+f_p^2)^2~~~~~~~~~~~~~~~~~~~~~~~~~~~~~~~~~\\
{\sf and}~~~~~~~~~~~~~~~~~~~~~&~~~~~~~~~~~~~~~~~~~~~~~~~~~~~~~~~~
~~~~~~~~~~~~~~~~~~~~~~~~~~~(2.55)\\
&{\cal P}\rightarrow \tau^2f_\epsilon^4\mu^2.
\end{array}
$$
If in addition $\tau\rightarrow 0$, then we have
$$
m_L^2\rightarrow m^2,~~~m_l^2\rightarrow \mu^2~~~{\sf
and}~~~m_s^2\rightarrow \tau^2 f_\epsilon^4.\eqno(2.56)
$$

\newpage

\section*{\Large \sf 3. Perturbative Solution and Jarlskog Invariant}

\noindent{\bf 3.1 Perturbation Series}\\

In this and the following sections we return to the mass Hamiltonian
(1.3) for quarks and calculate its eigenstates by using $G\gamma_4$
as the zeroth order Hamiltonian and $iF\gamma_4\gamma_5$ as the
perturbation. From the discussions given in the last section, we see
that this is identical to the problem of finding the eigenstates of
${\cal N}$ regarding $\tau$ as the small parameter. Using
(2.23)-(2.24), we may write
$$
{\cal N} = \left(
\begin{array}{ccc}
0 & 0 & 0\\
0 &\mu^2 & 0\\
0 & 0 & m^2
\end{array}
\right)+{\cal N}_1+O(\tau^2)\eqno(3.1)
$$
with
$$
{\cal N}_1 = \tau \left(
\begin{array}{ccc}
0 & -i\mu f_\epsilon f_p & -im f_P f_\epsilon\\
i\mu f_\epsilon f_p &0 & -i(m-\mu)f_p f_P\\
im f_P f_\epsilon & i(m-\mu)f_p f_P & 0
\end{array}
\right).\eqno(3.2)
$$
To first order in ${\cal N}_1$, the eigenstates of ${\cal N}$ can be
readily obtained. For applications to physical quarks, we need only
to identify that $f_\epsilon$, $f_p$ and $f_P$ are replaced by
$$
(f_\epsilon)_{\uparrow/\downarrow}=\tilde{f}\epsilon_{\uparrow/\downarrow},
$$
$$
(f_p)_{\uparrow/\downarrow}=\tilde{f}p_{\uparrow/\downarrow}\eqno(3.3)
$$
and
$$
(f_P)_{\uparrow/\downarrow}=\tilde{f}P_{\uparrow/\downarrow}.
$$
Likewise, neglecting $O(\tau^2)$ corrections, we can relate
$\mu^2$ and $m^2$ to (quark~mass)$^2$ by
$$
\mu^2_\uparrow = m_c^2,~~{\sf ~~~}~~m_\uparrow^2=m_t^2
$$
$$
\mu^2_\downarrow = m_s^2,~~{\sf
~~~}~~m_\downarrow^2=m_b^2\eqno(3.4)
$$
and set
$$
\tau=\tau_q.\eqno(3.5)
$$
Thus, to $O(\tau_q)$, in the $\uparrow$ sector the state vectors of
$u,~c,~t$ are related to those of $\epsilon_\uparrow,~p_\uparrow$
and $P_\uparrow$ by
$$
\left(\begin{array}{l}
u\\
c\\
t
\end{array}
\right)=\left(
\begin{array}{ccc}
(\epsilon_\uparrow|u)& (p_\uparrow|u) & (P_\uparrow|u)\\
(\epsilon_\uparrow|c) &(p_\uparrow|c)&(P_\uparrow|c)\\
(\epsilon_\uparrow|t) &(p_\uparrow|t)&(P_\uparrow|t)
\end{array}\right)\left(\begin{array}{l}
\epsilon_\uparrow\\
p_\uparrow\\
P_\uparrow
\end{array}
\right)\eqno(3.6)
$$
with
$$
(\epsilon_\uparrow|u)=1+O(\tau_q^2),~~~~~~~~~~~~~~\eqno(3.7)
$$
$$
(\epsilon_\uparrow|c)=-i\frac{\tau_q}{m_c}(f_\epsilon
f_p)_\uparrow+O(\tau_q^2),\eqno(3.8)
$$
$$
(\epsilon_\uparrow|t)=-i\frac{\tau_q}{m_t}(f_\epsilon
f_P)_\uparrow+O(\tau_q^2),\eqno(3.9)
$$

$$
(p_\uparrow|u)=-i\frac{\tau_q}{m_c}(f_\epsilon
f_p)_\uparrow+O(\tau_q^2),\eqno(3.10)
$$
$$
(p_\uparrow|c)=1+O(\tau_q^2),~~~~~~~~~~~~~~\eqno(3.11)
$$
$$
~~~~~~~~(p_\uparrow|t)=-i\frac{\tau_q}{m_t+m_c}(f_p
f_P)_\uparrow+O(\tau_q^2),\eqno(3.12)
$$

$$
~(P_\uparrow|u)=-i\frac{\tau_q}{m_t}(f_\epsilon
f_P)_\uparrow+O(\tau_q^2),\eqno(3.13)
$$
$$
~~~~~~~~(P_\uparrow|c)=-i\frac{\tau_q}{m_t+m_c}(f_p
f_P)_\uparrow+O(\tau_q^2)\eqno(3.14)
$$
and
$$
(P_\uparrow|t)=1+O(\tau_q^2).~~~~~~~~~~~~~\eqno(3.15)
$$
Likewise, for the $\downarrow$ sector, we may write
$$
\left(\begin{array}{l}
d\\
s\\
b
\end{array}
\right)=\left(
\begin{array}{ccc}
(\epsilon_\downarrow|d)& (p_\downarrow|d) &(P_\downarrow|d)\\
(\epsilon_\downarrow|s) &(p_\downarrow|s)&(P_\downarrow|s)\\
(\epsilon_\downarrow|b) &(p_\downarrow|b)&(P_\downarrow|b)
\end{array}\right)\left(\begin{array}{l}
\epsilon_\downarrow\\
p_\downarrow\\
P_\downarrow
\end{array}
\right)\eqno(3.16)
$$
with
$$
(\epsilon_\downarrow|d)=1+O(\tau_q^2),~~~~~~~~~~~~~~\eqno(3.17)
$$
$$
(\epsilon_\downarrow|s)=-i\frac{\tau_q}{m_s}(f_\epsilon
f_p)_\downarrow+O(\tau_q^2),\eqno(3.18)
$$
$$
(\epsilon_\downarrow|b)=-i\frac{\tau_q}{m_b}(f_\epsilon
f_P)_\downarrow+O(\tau_q^2),\eqno(3.19)
$$

$$
(p_\downarrow|d)=-i\frac{\tau_q}{m_s}(f_\epsilon
f_p)_\downarrow+O(\tau_q^2),\eqno(3.20)
$$
$$
(p_\downarrow|s)=1+O(\tau_q^2),~~~~~~~~~~~~~~\eqno(3.21)
$$
$$
~~~~~~~~(p_\downarrow|b)=-i\frac{\tau_q}{m_b+m_s}(f_p
f_P)_\downarrow+O(\tau_q^2),\eqno(3.22)
$$

$$
~(P_\downarrow|d)=-i\frac{\tau_q}{m_b}(f_\epsilon
f_P)_\downarrow+O(\tau_q^2),\eqno(3.23)
$$
$$
~~~~~~~~(P_\downarrow|s)=-i\frac{\tau_q}{m_b+m_s}(f_p
f_P)_\downarrow+O(\tau_q^2)\eqno(3.24)
$$
and
$$
(P_\downarrow|b)=1+O(\tau_q^2).~~~~~~~~~~~~~\eqno(3.25)
$$

\noindent{\bf 3.2 Jarlskog Invariant}\\

Write the CKM matrix as
$$
U_{CKM} = \left(
\begin{array}{ccc}
U_{11}& U_{12} &U_{13}\\
U_{21}& U_{22} &U_{23}\\
U_{31}& U_{32} &U_{33}
\end{array}\right).\eqno(3.26)
$$
Following Jarlskog[3], we introduce
$$
S_1=U_{11}^*U_{12},~~~S_2=U_{21}^*U_{22},~~~S_3=U_{31}^*U_{32}\eqno(3.27)
$$
and define
$$
{\cal J}=ImS_1^*S_2.\eqno(3.28)
$$
By using (3.27) we see that
$$
{\cal
J}=Im\bigg[(U_{11}U_{22})(U_{12}^*U_{21}^*)\bigg],\eqno(3.29)
$$
Because of unitarity of the CKM matrix,
$$
S_1+S_2+S_3=0.\eqno(3.30)
$$
Therefore, ${\cal J}$ is equal to twice the area of the triangle
whose sides are $S_1,~S_2$ and $S_3$. Furthermore, from the explicit
form of ${\cal J}$ given by (3.29), we see that ${\cal J}$ is
symmetric with respect to the interchange between the row and column
indices of the CKM matrix. It follows then in deriving ${\cal J}$,
we may use the elements of any two columns and of any two rows of
the CKM matrix.

It is convenient to denote $(U_{CKM})_0$ of (1.21) simply as $V$,
with
$$
V\equiv (U_{CKM})_0 = \left(
\begin{array}{ccc}
V_{11}& V_{12} &V_{13}\\
V_{21}& V_{22} &V_{23}\\
V_{31}& V_{32} &V_{33}
\end{array}\right).\eqno(3.31)
$$
In terms of the state vectors
$\epsilon_\uparrow,~p_\uparrow.~P_\uparrow$ and
$\epsilon_\downarrow,~p_\downarrow,~P_\downarrow$ of
(1.13)-(1.18), we can also write $V$ as
$$
V=\left(
\begin{array}{ccc}
(\epsilon_\uparrow|\epsilon_\downarrow)&
(\epsilon_\uparrow|p_\downarrow) &
(\epsilon_\uparrow|P_\downarrow)\\
(p_\uparrow|\epsilon_\downarrow) &(p_\uparrow|p_\downarrow)&
(p_\uparrow|P_\downarrow)\\
(P_\uparrow|\epsilon_\downarrow)
&(P_\uparrow|p_\downarrow)&(P_\uparrow|P_\downarrow)
\end{array}\right).\eqno(3.32)
$$
Likewise, the CKM matrix is given by
$$
U_{CKM}= \left(
\begin{array}{ccc}
(u|d)& (u|s) & (u|b)\\
(c|d) &(c|s)&(c|b)\\
(t|d) &(t|s)&(t|b)
\end{array}\right)
$$
$$
~~~~=V+i\tau_qW+O(\tau_q^2)\eqno(3.33)
$$
where the matrix elements of $W$ are derived from (3.7)-(3.15) and
(3.17)-(3.25). Using the perturbative solution of Sec. 3.1, we can
readily express the matrix elements of $U_{CKM}$ in terms of the
corresponding ones of $V$. The Jarlskog invariant can then be
evaluated by using (3.29).

As will be shown in Appendix B, the result, accurate to the first
power of $\tau_q$, is
$$
{\cal J}=\tau_q\bigg[\frac{(f_\epsilon f_p)_\downarrow}{m_s}A_s+
\frac{(f_\epsilon f_P)_\downarrow}{m_b}A_b +\frac{(f_p
f_P)_\downarrow}{m_s+m_b}B_\downarrow
$$
$$
+ \frac{(f_\epsilon f_p)_\uparrow}{m_c}A_c+\frac{(f_\epsilon
f_P)_\uparrow}{m_t}A_t+ \frac{(f_p
f_P)_\uparrow}{m_c+m_t}B_\uparrow\bigg]\eqno(3.34)
$$
where
$$
A_s=-V_{13}V_{23}V_{33}\cong -2\cdot 10^{-4},\eqno(3.35)
$$
$$
A_b=-V_{12}V_{22}V_{32}\cong 8.8\cdot 10^{-3},~\eqno(3.36)
$$
$$
B_\downarrow=-V_{11}V_{21}V_{31}\cong 1.10\cdot
10^{-3},\eqno(3.37)
$$
$$
A_c=V_{31}V_{32}V_{33}\cong -2\cdot 10^{-4},~~\eqno(3.38)
$$
$$
A_t=V_{21}V_{22}V_{23}\cong -8.8\cdot 10^{-3}\eqno(3.39)
$$
and
$$
B_\uparrow=V_{11}V_{12}V_{13}\cong 1.10\cdot 10^{-3}.\eqno(3.40)
$$

From the definition (3.29) and (3.34), these coefficients
$A_s,~\cdots,~B_\uparrow$ are all products of four factors of
$V_{ij}$. As will be shown in Appendix B, because $V$ is a real
orthogonal matrix, these quartic products can all be reduced to
triple products given by (3.35)-(3.40). Since $m_c>>m_s$ and
$m_t>>m_b$, we can, as an approximation, neglect the terms related
to the up sector in (3.34).

\section*{\Large \sf 4. Determination of $\tau_q$ and $f$}

\noindent{\bf 4.1 A Special coordinate system}\\

For the $\uparrow$ quarks, the parameters $\lambda_1,~\lambda_2$ and
$\lambda_3$ in (2.25)-(2.29) are related to the quark masses by
$$
\lambda_1=m_u,~~\lambda_2=m_c~~{\sf and}~~\lambda_3=m_t.\eqno(4.1)
$$
Likewise, $f_\epsilon$ is
$$
(f_\epsilon)_\uparrow=\tilde{f}\epsilon_\uparrow\eqno(4.2)
$$
with $\epsilon_\uparrow$ given by (1.13) and $f$ the unit
directional vector of (1.24). We work to the lowest order in
$\tau_q$. From (2.26), setting $\lambda_1\lambda_2\lambda_3=m_u\mu
m$ we have
$$
m_u=\tau_q(\tilde{f}\epsilon_\uparrow)^2.\eqno(4.3)
$$
Likewise, for the $\downarrow$ quarks
$$
m_d=\tau_q(\tilde{f}\epsilon_\downarrow)^2.\eqno(4.4)
$$
It is convenient to introduce a special coordinate system in which
$$
\epsilon_\downarrow=\left(\begin{array}{l}
1\\
0\\
0
\end{array}
\right)~~{\sf and}~~ \epsilon_\uparrow=\left(\begin{array}{l}
~~\cos\theta_c\\
-\sin\theta_c\\
~~~~0
\end{array}
\right).\eqno(4.5)
$$
Since $p_\downarrow$ and $P_\downarrow$ are both $\bot~
\epsilon_\downarrow$, we may write
$$
p_\downarrow=\left(\begin{array}{l}
~~~~0\\
-\cos\gamma\\
~~\sin\gamma
\end{array}
\right)~~{\sf and}~~ P_\downarrow=\left(\begin{array}{l}
~~~~0\\
-\sin\gamma\\
-\cos\gamma
\end{array}
\right).\eqno(4.6)
$$
Furthermore, we shall set the zeroth order CKM matrix
$(U_{CKM})_0$ of (1.21) to be
$$
(U_{CKM})_0=\left(
\begin{array}{ccc}
(\epsilon_\uparrow|\epsilon_\downarrow)&
(\epsilon_\uparrow|p_\downarrow) &
(\epsilon_\uparrow|P_\downarrow)\\
(p_\uparrow|\epsilon_\downarrow) &(p_\uparrow|p_\downarrow)&(p_\uparrow|P_\downarrow)\\
(P_\uparrow|\epsilon_\downarrow)
&(P_\uparrow|p_\downarrow)&(P_\uparrow|P_\downarrow)
\end{array}\right)
$$
$$
=\left(
\begin{array}{ccc}
~~.974& ~~~~.227 & ~~5\cdot 10^{-3}\\
-.227& ~~~~.973 & ~.04\\
5\cdot 10^{-3}& -.04 & ~~~.999
\end{array}\right)+O(1\cdot 10^{-3}).\eqno(4.7)
$$
Thus, with the same accuracy of $O(10^{-3})$, the Cabibbo angle
$\theta_c$ is given by
$$
\cos \theta_c=.974~~{\sf and}~~\sin\theta_c=.227.\eqno(4.8)
$$
Likewise, from $(\epsilon_\uparrow|P_\downarrow)=5\cdot 10^{-3}$ in
(4.7), in accordance with (4.5), (4.6) and
$$
(\epsilon_\uparrow|P_\downarrow)=\sin\theta_c\sin\gamma,\eqno(4.9)
$$
we find
$$
\sin\gamma=2.2\cdot 10^{-2},\eqno(4.10)
$$
which together with (4.5) and (4.6) give the coordinate system
defined by $(\epsilon_\downarrow,~p_\downarrow,P_\downarrow)$.
Eq.(4.7) then, in turn, determines the corresponding coordinate
system  $(\epsilon_\uparrow,~p_\uparrow,P_\uparrow)$.

Next, we shall determine the parameters $\tau_q$ and the directional
angles $\alpha$ and $\beta$ of the unit vector
$$
f= \left(
\begin{array}{l}
\sin\alpha\cos\beta\\
\sin\alpha\sin\beta\\
\cos\alpha
\end{array}
\right)\eqno(4.11)
$$
in the coordinate system defined by (4.5)-(4.6).\\

\newpage

\noindent{\bf 4.2 Determination of $\beta$}\\

From (4.5) and (4.11), we have
$$
\tilde{f}\epsilon_\downarrow=\sin\alpha\cos\beta\eqno(4.12)
$$
and
$$
\tilde{f}\epsilon_\uparrow=\sin\alpha\cos(\beta+\theta_c).\eqno(4.13)
$$
Thus, on account of (4.3) and (4.4),
$$
\frac{\cos^2(\beta+\theta_c)}{\cos^2\beta}=\frac{m_u}{m_d}\eqno(4.14)
$$
and therefore
$$
\frac{\cos(\beta+\theta_c)}{\cos\beta}=
\pm\bigg(\frac{m_u}{m_d}\bigg)^{\frac{1}{2}}.\eqno(4.15)
$$
assuming
$$
\frac{m_u}{m_d}\cong \frac{1}{2},\eqno(4.16)
$$
we find two solutions for $\beta$:
$$
\beta\cong 48^0~50'\eqno(4.17)
$$
or
$$
\beta\cong 82^0~20'.\eqno(4.18)
$$

\noindent{\bf 4.3 Determination of $\alpha$ and $\tau_q$}\\

We shall first determine the parameter $\alpha$ by using the
Jarlskog invariant
$$
{\cal J}=3.08\cdot 10^{-5}.\eqno(4.19)
$$
Define
$$
F=10^2{\cal J}m_b/\tau_q.\eqno(4.20)
$$
From (4.4), (4.5) and (4.11), we have
$$
m_d=\tau_q\sin^2\alpha\cos^2\beta\eqno(4.21)
$$
and therefore
$$
F=10^2{\cal J}(m_b/m_d)\sin^2\alpha\cos^2\beta.\eqno(4.22)
$$
For definiteness, we shall set the various quark masses as
$$
m_d \cong 5MeV,~~~~~~~m_u\cong 2.5MeV
$$
$$
m_s\cong 95MeV,~~~~~~~m_c\cong 1.25GeV
$$
$$
m_b\cong 4.2GeV~~{\sf and}~~m_t\cong 175GeV,\eqno(4.23)
$$
consistent with the Particle Data Group values[7]. Thus, (4.22)
becomes
$$
F(\alpha,~\beta)\cong 2.6\sin^2\alpha\cos^2\beta,\eqno(4.24)
$$

On the other hand, from (3.34) and by using the numerical values for
$A_s,~A_b,~\cdots,~B_\uparrow$ of (3.35)-(3.40) together with the
various quark masses given above, the same function
$F(\alpha,~\beta)$ is also
$$
F(\alpha,~\beta)\cong -0.88(f_\epsilon
f_p)_\downarrow~~-0.067~(f_\epsilon f_p)_\uparrow
$$
$$
~~~~~~~~~~~~~ +0.88(f_\epsilon f_P)_\downarrow~~-0.021~(f_\epsilon
f_P)_\uparrow
$$
$$
~~~~~~~~~~~~~~ +0.11(f_p f_P)_\downarrow~~+0.0026(f_p
f_P)_\uparrow.\eqno(4.25)
$$
As an approximation, we may neglect the contributions of the
$\uparrow$ sector. Combining (4.24) with (4.25), we find
$$
2.6 \sin^2\alpha\cos^2\beta\cong 0.88
f_{\epsilon_\downarrow}[f_{P_\downarrow}-f_{p_\downarrow}]
+0.11(f_pf_P)_\downarrow\eqno(4.26)
$$
with
$$
f_{\epsilon_\downarrow}=\tilde{f}\epsilon_\downarrow\eqno(4.27)
$$
given by (4.12),
$$
f_{p_\downarrow}=\tilde{f}p_\downarrow=
-\sin\alpha\sin\beta\cos\gamma+\cos\alpha\sin\gamma\eqno(4.28)
$$
and
$$
f_{P_\downarrow}=\tilde{f}P_\downarrow=
-\sin\alpha\sin\beta\sin\gamma-\cos\alpha\cos\gamma.\eqno(4.29)
$$
By using $\gamma$ from (4.10),
$$
\beta\cong48^0~50'
$$
from (4.17), we find
$$
\alpha\cong -36^0~10'.\eqno(4.30)
$$
From (4.4) and (4.12), we have
$$
\tau_q=m_d/(\sin^2\alpha\cos^2\beta).\eqno(4.31)
$$
The above values of $\alpha,~\beta$ and $m_d\cong 5MeV$ give
$$
\tau_q\cong 33MeV.\eqno(4.32)
$$

On the other hand, the alternative solution $\beta\cong 82^0~20'$ of
(4.18) leads to a much larger value $\tau_q\sim 5.5GeV$. Such a
large value invalidates the small $\tau_q$ approximation. Thus, we
focus only on the solution (4.32) in this paper.

\section*{\Large \sf 5. Applications to Leptons}

The mapping matrix for leptons presents us with a different
quantitative picture from the CKM matrix for quarks.

In the CKM matrix, the only off-diagonal elements of a relatively
significant size are those associated with the Cabibbo angle
$\theta_c$, which is also not large. In the lepton mapping matrix
$U_l$, except for its $T$-violating element ($(U_l)_{13}$ in the
standard form), all matrix elements are not small. Furthermore,
within the present $\sim 1\sigma$ accuracy, $U_l$ is given by the
Harrison-Perkins-Scott (HPS) form[8]
$$
(U_l)_0=\left(
\begin{array}{rrc}
\sqrt{\frac{2}{3}} & \sqrt{\frac{1}{3}} & 0\\
-\sqrt{\frac{1}{6}} & \sqrt{\frac{1}{3}} & \sqrt{\frac{1}{2}} \\
\sqrt{\frac{1}{6}} & -\sqrt{\frac{1}{3}} & \sqrt{\frac{1}{2}}
\end{array}\right). \eqno(5.1)
$$
This leads us to propose the following simple Timeon model for
leptons.\\

\noindent{\bf 5.1 Mass Matrices for Leptons}\\

As in (1.1)-(1.2), we define $l_i(\uparrow)$ and $l_i(\downarrow)$
to be the lepton states "diagonal" in $W^\pm$ transitions, so that
$$
\begin{array}{ll}
&l_i(\uparrow)\rightleftharpoons l_i(\downarrow) + W^+\\
{\sf and}~~~~~~~~~~~~~~~~~~~~~&~~~~~~~~~~~~~~~~~~~~~~~~~~~~~~~~~~
~~~~~~~~~~~~~~~~~~~~~~~~~~~(5.2)\\
&l_i(\downarrow)\rightleftharpoons l_i(\uparrow) + W^-.
\end{array}
$$
with $i=1,~2$ and $3$. Their electric charges in units of $e$ are
$0$ for $l_i(\uparrow)$ and $-1$ for $l_i(\downarrow)$. However,
these $l_i(\uparrow)$ and $l_i(\downarrow)$ are not the mass
eigenstates $\nu_1,~\nu_2,~\nu_3$ and $e,~\mu,~\tau$. Their mass
Hamiltonians $H_\uparrow$ and $H_\downarrow$ are given by
$$
H_{\uparrow/\downarrow}=\bigg(l_1^\dag,~
l_2^\dag,~l_3^\dag\bigg)_{\uparrow/\downarrow} (G(l) \gamma_4 +
iF(l)\gamma_4 \gamma_5)_{\uparrow/\downarrow}\left(
\begin{array}{r}
l_1\\
l_2\\
l_3
\end{array}\right)_{\uparrow/\downarrow}.\eqno(5.3)
$$
We choose $G_\uparrow$ and $G_\downarrow$ so that if
$F_{\uparrow/\downarrow}$ were to vanish, the lepton mapping
matrix would be (5.1) exactly and, in addition, the lowest lepton
mass in either $\uparrow$ or $\downarrow$ would be zero. We
further simplify our model by assuming that the timeon field does
not couple to neutrinos; i.e.,
$$
F(l)_\uparrow=0.\eqno(5.4)
$$
 Thus, the lightest neutrino is massless. All departures of the
 lepton mapping matrix from the HPS form (5.1) stem from the single
 timeon term
 $$
 F(l)_\downarrow=\tau_l v\tilde{v}\eqno(5.5)
 $$
for the charged leptons, with $v$ a real $3$ dimensional unit vector
like $f$ of (1.24) for quarks. The same timeon term (5.5) also gives
the electron mass. Thus, all departures from (5.1) would come from
the charged leptons. A consequence of the assumption
$F(l)_\uparrow=0$ is that the predictions of the model will now
depend on the choice (formerly arbitrary) of the "hidden" bases
$(l_1,~l_2,~l_3)_{\uparrow/\downarrow}$ of (5.2). As illustrated in
Figure 2, we have chosen this basis to recognize the factorization
of (5.1) with
$$
(U_l)_0\equiv (V_\downarrow)_0^\dag(V_\uparrow)_0\eqno(5.6)
$$
in which
$$
(V_\downarrow)_0=\left(
\begin{array}{crr}
1 & 0 & 0 \\
0 & \sqrt{\frac{1}{2}} & -\sqrt{\frac{1}{2}}\\
0 & \sqrt{\frac{1}{2}} & \sqrt{\frac{1}{2}}
\end{array}\right) \eqno(5.7)
$$
and
$$
(V_\uparrow)_0=\left(
\begin{array}{rcc}
\sqrt{\frac{2}{3}} & \sqrt{\frac{1}{3}} & 0\\
-\sqrt{\frac{1}{3}} & \sqrt{\frac{2}{3}} & 0 \\
0 & 0 & 1
\end{array}\right). \eqno(5.8)
$$
It appears natural to associate the left-hand factor in the
product (5.6) with the charged leptons and the right-hand factor
with the neutrinos; i.e., through $(V_\downarrow)_0$ of (5.7), we
identify
$$
(l_1)_\downarrow=e_0\eqno(5.9)
$$
and through $(V_\uparrow)_0$ of (5.8),
$$
(l_3)_\uparrow=\nu_3.\eqno(5.10)
$$
In (5.9), $e_0$ denotes the zeroth order electron state (i.e.,
without its timeon correction). It is convenient to consider (5.6)
as a product of two consecutive rigid body rotations with $e_0$ and
$\nu_3$ as their respective fixed axes of rotations of the same
rigid body. In this picture, we may identify
$$
l_i(\uparrow)=l_i(\downarrow).\eqno(5.11)
$$
A comparison between (5.9) and (5.10) suggests that like $e_0$,
$\nu_3$ is also massless. (There is an alternative possibility
with $\nu_1$ massless, as will also be discussed below.)\\

\noindent{\bf 5.2 Analysis of Mass Matrices }\\

We begin with $H_\downarrow$ of (5.3) for the charged leptons by
assuming
$$
G(l)_\downarrow=\left(
\begin{array}{ccc}
0 & 0 & 0\\
0 & a_\downarrow + b_\downarrow & -a_\downarrow\\
0 & -a_\downarrow & a_\downarrow + b_\downarrow
\end{array}\right), \eqno(5.12)
$$
with $a_\downarrow$, $b_\downarrow$ both positive. Using (5.7) we
can diagonalize $G(l)_\downarrow$ through
$$
(V_\downarrow)_0^\dag G(l)_\downarrow(V_\downarrow)_0=\left(
\begin{array}{ccc}
0 & 0 & 0\\
0 & b_\downarrow & 0 \\
0 & 0 & 2a_\downarrow + b_\downarrow
\end{array}\right) \eqno(5.13)
$$
and leads to the zeroth order mass of $e$ to be $0$, those of mu
and tau to be
$$
\begin{array}{ll}
&\mu=b_\downarrow\\
{\sf and}~~~~~~~~~~~~~~~~~~~~~&~~~~~~~~~~~~~~~~~~~~~~~~~~~~~~~~~~
~~~~~~~~~~~~~~~~~~~~~~~~~~~(5.14)\\
&m=2a_\downarrow+b_\downarrow.
\end{array}
$$
The physical masses of $e$, $\mu$ and $\tau$ can then be obtained
by using (2.23)-(2.29), with the parameters $\tau$ and $f$
replaced by $\tau_l$ and $v$ of the leptonic timeon term (5.5).

In the $\uparrow$ sector, in accordance with (5.4) the neutrino
masses are only due to the $G(l)_\uparrow$ term, with the lowest
neutrino mass zero. There are two such possibilities, of which the
first is

\noindent (i)
$$
G(l)_\uparrow=\left(
\begin{array}{ccc}
\frac{1}{2}a_\uparrow+b_\uparrow & \sqrt{\frac{1}{2}}a_\uparrow & 0\\
\sqrt{\frac{1}{2}}a_\uparrow & a_\uparrow+b_\uparrow & 0\\
0 & 0 & 0
\end{array}\right), \eqno(5.15)
$$
where $a_\uparrow$ and $b_\uparrow$ are both positive. The masses
of $\nu_1$, $\nu_2$ and $\nu_3$ are given by
$$
m_1=b_\uparrow,~~~m_2=\frac{3}{2}a_\uparrow+b_\uparrow\eqno(5.16)
$$
and
$$
m_3=0.\eqno(5.17)
$$
In this case, we would have from the present experimental
data[9,10]
$$
\begin{array}{ll}
&\frac{1}{2}(m_1^2+m_2^2)\cong 2.39\times 10^{-3}eV^2\\
{\sf and}~~~~~~~~~~~~~~~~~~~~~&~~~~~~~~~~~~~~~~~~~~~~~~~~~~~~~~~~
~~~~~~~~~~~~~~~~~~~~~~~~~~~(5.18)\\
&m_2^2-m_1^2 \cong 7.67\times 10^{-5}eV^2.
\end{array}
$$
\noindent (ii) The second possibility is,
$$
G(l)_\uparrow=\left(
\begin{array}{rrc}
\frac{1}{2}a_\uparrow & \sqrt{\frac{1}{2}}a_\uparrow & 0\\
\sqrt{\frac{1}{2}}a_\uparrow & a_\uparrow & 0\\
0 & 0 & b_\uparrow
\end{array}\right), \eqno(5.19)
$$
which leads to
$$
\begin{array}{ll}
&m_1=0,\\
{\sf and}~~~~~~~~~~~~~~~~~~~~~&~~~~~~~~~~~~~~~~~~~~~~~~~~~~~~~~~~
~~~~~~~~~~~~~~~~~~~~~~~~~~~(5.20)\\
&m_2=\frac{3}{2}a_\uparrow,~m_3=b_\uparrow.
\end{array}
$$
In this case, instead of (5.18), we have
$$
\begin{array}{ll}
&m_2^2\cong 7.67\times 10^{-5}eV^2\\
{\sf and}~~~~~~~~~~~~~~~~~~~~~&~~~~~~~~~~~~~~~~~~~~~~~~~~~~~~~~~~
~~~~~~~~~~~~~~~~~~~~~~~~~~~(5.21)\\
&m_3^2 \cong 2.43\times 10^{-3}eV^2.
\end{array}
$$

In either case, (i) or (ii) the mass matrix (5.15) or (5.19) can
be diagonalized with the same real unitary matrix $
(V_\uparrow)_0$ of (5.8) so that
$$
(V_\uparrow)_0^\dag G(l)_\uparrow(V_\uparrow)_0=\left(
\begin{array}{rcc}
m_1 & 0 & 0\\
0 & m_2 & 0 \\
0 & 0 & m_3
\end{array}\right), \eqno(5.22)
$$
and the zeroth order neutrino mapping matrix (5.6) is the HPS form
(5.1).

In Fig.~2, we give a graphic illustration of these two rotations
$(V_\downarrow)_0$ and $(V_\uparrow)_0$. In either (i) or (ii),
the parameters in the neutrino sector are determined. The
remaining parameters are
$$
\mu,~~m\eqno(5.23)
$$
the two mass parameters of (5.14) for the charged leptons, and the
three parameters
$$
\tau_l~~{\sf and~two~angular~variables}\eqno(5.24)
$$
that characterize the unit vector $v$ of the timeon term (5.5). On
the other hand, these five parameters in (5.23)-(5.24) should
account for seven observables: the three charged lepton masses
$$
m_e,~~m_\mu,~~m_\tau\eqno(5.25)
$$
and the four parameters
$$
\theta_{12},~~\theta_{23},~~\theta_{31}\eqno(5.26)
$$
and
$$
{\sf the~Jarlskog~invariant}~~{\cal J}_l\eqno(5.27)
$$
of the lepton mapping matrix. Hence, the model defined in this
section predicts two relations between these seven observables
(5.25)-(5.27), as we shall discuss. At present, the existing
knowledge of the neutrino mapping matrix agrees with the HPS form
(5.1) to $\approx 10\%$. Anticipating that future experiments may
improve the accuracy to $1\%$ level, we shall calculate the elements
of the lepton mapping matrix to all orders in $\tau_l/\mu$, first
order in $\tau_l/m$, but neglecting all $1/m^2$ corrections.\\

\noindent{\bf 5.3 Statevectors $e$, $\mu$ and $\tau$}\\

Consider the $l_i(\downarrow)$ sector. From (5.13)-(5.14), we have
$$
(V_\downarrow)_0^\dag G(l)_\downarrow(V_\downarrow)_0=\left(
\begin{array}{ccc}
0 & 0 & 0\\
0 & \mu & 0 \\
0 & 0 & m
\end{array}\right). \eqno(5.28)
$$
As in (2.12) and (2.14), we can express $G(l)_\downarrow$ in terms
of its eigenvalues $0$, $\mu$, $m$ and their corresponding
eigenvectors $\epsilon$, $p$, $P$ as
$$
G(l)_\downarrow=\mu p\tilde{p} +mP\tilde{P}. \eqno(5.29)
$$
Under the same transformation $(V_\downarrow)_0$, the $T$ odd
timeon term $F(l)_\downarrow$ of (5.5) becomes
$$
(V_\downarrow)_0^\dag F(l)_\downarrow (V_\downarrow)_0=\tau_l
f'\tilde{f'}\eqno(5.30)
$$
with $f'$ related to the unit vector $v$ of (5.5) by
$$
f'=(V_\downarrow)_0^\dag v.\eqno(5.31)
$$
As in (2.19), write
$$
f'=\left(\begin{array}{l}
f_\epsilon\\
f_p\\
f_P
\end{array}
\right),\eqno(5.32)
$$
where in place of (2.20) we have
$$
f_\epsilon
=\tilde{\epsilon}v,~~f_p=\tilde{p}v,~~~f_P=\tilde{P}v.\eqno(5.33)
$$
Equate the matrices ${\cal G}$ and ${\cal F}$ of (2.1) with
$G(l)_\downarrow$ and $F(l)_\downarrow$, the matrix ${\cal N}$ of
(2.24) becomes
$$
{\cal N}=(V_\downarrow)_0^\dag {\cal M}^2
(V_\downarrow)_0\eqno(5.34)
$$
with ${\cal M}^2$ of (2.7) given now by
$$
{\cal M}^2=[G(l)_\downarrow-iF(l)_\downarrow]
[G(l)_\downarrow+iF(l)_\downarrow].\eqno(5.35)
$$

For applications to charged leptons, we identify the subscripts
$s$, $l$ and $L$ of (2.32) in Section 2.2 with $e$, $\mu$ and
$\tau$. Thus, (2.31) and (2.32) become
$$
E_i=\lambda_i^2=m_i^2\eqno(5.36)
$$
and
$$
i=e,~\mu~~{\sf and}~~\tau.\eqno(5.37)
$$
These lepton masses and their corresponding eigenvectors $\psi_e$,
$\psi_\mu$ and $\psi_\tau$ can be readily obtained by using results
derived in Sections 2.2 and 2.3, as we shall see.

For $i=e$ and $\mu$, the eigenfunction $\psi_i$ can be written as,
in accordance with (2.38),
$$
\psi_e=\left(
\begin{array}{c}
\phi_e\\
c_e
\end{array}\right)~~{\sf and}~~\psi_\mu=\left(
\begin{array}{c}
\phi_\mu\\
c_\mu
\end{array}\right).\eqno(5.38)
$$
By neglecting $n^{-2}=O(m^{-2})$, we can approximate (2.44)as
$$
(h-\chi\chi^\dag)\phi_i=E_i\phi_i\eqno(5.39)
$$
with
$$
E_i=m_e^2~~{\sf or}~~m_\mu^2.\eqno(5.40)
$$
To the same approximation, (2.40) gives
$$
c_i=-\chi^\dag\phi_i/n.\eqno(5.41)
$$
In terms of the Pauli spin matrices $\sigma_i$, write
$$
h-\chi\chi^\dag=a+b_1\sigma_1+b_2\sigma_2+b_3\sigma_3\eqno(5.42)
$$
where, neglecting $O(m^{-2})$,
$$
a~=\frac{1}{2}[\mu^2+\tau_l^2(f_\epsilon^2+f_p^2)^2+2\frac{\mu}{m}\tau_l^2f_p^2f_P^2]\eqno(5.43)
$$
and
$$
\begin{array}{rll}
~~~~~~~~~~~~~~~~~~~~~~b_1=&\tau_l^2(f_\epsilon^2+f_p^2+\frac{\mu}{m}f_P^2)f_\epsilon f_p,&\\
b_2=&\tau_l\mu f_\epsilon f_p,~~~~~~~~~~~~~~~~~~~~~~~~~~~~~~~~~~~~~~~~~~~~~~&(5.44)\\
b_3=&-\frac{1}{2}\mu^2+\frac{1}{2}\tau_l^2[(f_\epsilon^4-f_p^4)-2\frac{\mu}{m}f_p^2f_P^2].&
\end{array}
$$
The eigenvalues of (5.42) are
$$
\begin{array}{ll}
&m_\mu^2=a+b\\
{\sf and}~~~~~~~~~~~~~~~~~~~~~&~~~~~~~~~~~~~~~~~~~~~~~~~~~~~~~~~~
~~~~~~~~~~~~~~~~~~~~~~~~~~~(5.45)\\
&m_e^2 =a-b
\end{array}
$$
with
$$
b=(b_1^2+b_2^2+b_3^2)^{\frac{1}{2}}.~~~~~~~~~~\eqno(5.46)
$$

In the charged lepton sector, it is convenient to introduce the
polar coordinates $\alpha_l$ and $\beta_l$ through
$$
b_1=b\sin \alpha_l\cos\beta_l,~~b_2=b\sin\alpha_l\sin\beta_l~~{\sf
and}~~b_3=b\cos\alpha_l.\eqno(5.47)
$$


\noindent Neglecting $O(m^{-2})$, we find the components of the
eigenfunctions $\psi_e$ and $\psi_\mu$ in ( 5.38) to be
$$
\phi_e=N_e\left(
\begin{array}{c}
\sin\frac{1}{2}\alpha_l\\
-e^{i\beta_l}\cos\frac{1}{2}\alpha_l
\end{array}\right)
$$
$$
{\sf and}~~~~~~~~~~~~~~~~~~~~~~~~~~~~~~~~~~~~~~~~~~~~~~~~~~~~~~~
~~~~~~~~~~~~~~~~~~~~~~~~~~~(5.48)
$$
$$
\phi_\mu=N_\mu\left(
\begin{array}{c}
e^{-i\beta_l}\cos\frac{1}{2}\alpha_l\\
\sin\frac{1}{2}\alpha_l
\end{array}\right).
$$
There are arbitrary phase factors in  $N_e$ and $N_\mu$, which will
be discussed in Appendix C. Here we simply set these normalization
factors to be
$$
N_e=N_\mu=1+O(m^{-2}).\eqno(5.49)
$$
Combining (5.38) with (5.48), we have
$$
\psi_e=\left(
\begin{array}{c}
<\epsilon_\downarrow|e>\\
<p_\downarrow|e>\\
<P_\downarrow|e>
\end{array}\right)=
\left(
\begin{array}{c}
\sin\frac{1}{2}\alpha_l\\
-e^{i\beta_l}\cos\frac{1}{2}\alpha_l\\
c_e
\end{array}\right)\eqno(5.50)
$$
and
$$
\psi_\mu=\left(
\begin{array}{c}
<\epsilon_\downarrow|\mu>\\
<p_\downarrow|\mu>\\
<P_\downarrow|\mu>
\end{array}\right)=
\left(
\begin{array}{c}
e^{-i\beta_l}\cos\frac{1}{2}\alpha_l\\
\sin\frac{1}{2}\alpha_l\\
c_\mu
\end{array}\right)\eqno(5.51)
$$
in which
$$
c_e=-m^{-1}\chi^\dag\left(
\begin{array}{c}
\sin\frac{1}{2}\alpha_l\\
-e^{i\beta_l}\cos\frac{1}{2}\alpha_l
\end{array}\right),\eqno(5.52)
$$
$$
c_\mu=-m^{-1}\chi^\dag\left(
\begin{array}{c}
e^{-i\beta_l}\cos\frac{1}{2}\alpha_l\\
\sin\frac{1}{2}\alpha_l
\end{array}\right)
$$
and
$$
\chi\equiv\left(
\begin{array}{c}
\chi_\epsilon\\
\chi_p
\end{array}\right)\eqno(5.53)
$$
given by (2.36)-(2.37). Correspondingly, neglecting $O(m^{-2})$ we
find
$$
\psi_\tau=\left(
\begin{array}{c}
<\epsilon_\downarrow|\tau>\\
<p_\downarrow|\tau>\\
<P_\downarrow|\tau>
\end{array}\right)=
\left(
\begin{array}{c}
m^{-1}\chi_\epsilon\\
m^{-1}\chi_p\\
1
\end{array}\right).\eqno(5.54)
$$
Define the transformation matrix $W$ to be
$$
W=\left(
\begin{array}{ccc}
<e|\epsilon_\downarrow>&<\mu|\epsilon_\downarrow>&<\tau|\epsilon_\downarrow>\\
<e|p_\downarrow>&<\mu|p_\downarrow>&<\tau|p_\downarrow>\\
<e|P_\downarrow>&<\mu|P_\downarrow>&<\tau|P_\downarrow>
\end{array}\right)\eqno(5.55)
$$
with $\epsilon_\downarrow$,~$p_\downarrow$,~$P_\downarrow$ being
the same eigenvectors $\epsilon,~p,~P$ of (5.29). Thus,
$$
\left(
\begin{array}{c}
\epsilon\\
p\\
P
\end{array}\right)_\downarrow=W
\left(
\begin{array}{c}
e\\
\mu\\
\tau
\end{array}\right),
$$
$$
\eqno(5.56)
$$
$$
\left(
\begin{array}{c}
e\\
\mu\\
\tau
\end{array}\right)=W^\dag\left(
\begin{array}{c}
\epsilon\\
p\\
P
\end{array}\right)_\downarrow
$$
and the lepton mapping matrix is given by
$$
V_{l-{\sf map}}=W^\dag\cdot (U_l)_0\eqno(5.57)
$$
with $(U_l)_0$ given by the HPS form (5.1).

Combining these matrices, we derive the lepton mapping matrix
$V_{l-{\sf map}}$ given in Table~1. To compare with the
experimental neutrino mapping matrix, there is still a phase
convention which will be discussed in
Appendix C.\\

\noindent{\bf 5.4 Jarlskog Invariant}\\

The Jarlskog invariant  ${\cal J}_l$ can be calculated by using
(3.29) and replacing $U_{CKM}$ by $V_{l-{\sf map}}$ of Table~1. The
result is
$$
{\cal J}_l=\frac{N}{6(m_\mu^2-m_e^2)}[1+O(m^{-2})]\eqno(5.58)
$$
with
$$
N=\mu\tau_lf_\epsilon f_p+(\tau_l/m)f_\epsilon
f_P\bigg[\mu^2-\tau_l^2(f_\epsilon^2+f_p^2)
(f_\epsilon-f_p)(f_\epsilon+2f_p)\bigg] \eqno(5.59)
$$
In the limit $m\rightarrow \infty$, ${\cal J}_l$ becomes
$$
{\cal J}_l=[6(m_\mu^2-m_e^2)]^{-1}m_\mu\tau_l f_\epsilon
f_p[1+O(m^{-1})].\eqno(5.60)
$$
In the limit $\tau_l\rightarrow 0$, we have
$$
{\cal J}_l=\frac{1}{6}\tau_l (m_\mu^{-1}f_\epsilon
f_p+m_\tau^{-1}f_\epsilon f_P)+O(\tau_l^3).\eqno(5.61)
$$
The same result can also be derived by using the perturbative
expression (3.34) and replacing $V$ in (3.35)-(3.40) by $V_{l-{\sf
map}}$.\\

\noindent {\bf \underline{Remarks}}\\

As mentioned before, the leptonic timeon model predicts two
relations between the seven observables (5.25)-(5.27). To see how
a test can be made, we may take the five parameters in the theory
to be
$$
m,~\mu,~\tau_l,~f_\epsilon~~{\sf and}~~f_p\eqno(5.62)
$$
of (5.23)-(5.24), or the equivalent set
$$
m,~a,~b,~\alpha_l~~{\sf and}~~\beta_l\eqno(5.63)
$$
with $a,~b$ given by (5.43), (5.46) and $\alpha_l,~\beta_l$ by
(5.47). Using (2.50) and setting $m_L=m_\tau$, the mass of the
heavy lepton $\tau$, we have for the first parameter in (5.63)
$$
m^2=m_\tau^2+O(\tau_l^2).\eqno(5.64)
$$
Likewise, from (5.45)
$$
a=\frac{1}{2}(m_\mu^2+m_e^2)\eqno(5.65)
$$
and
$$
b=\frac{1}{2}(m_\mu^2-m_e^2).\eqno(5.66)
$$
The two remaining angular parameters $\alpha_l$ and $\beta_l$ in
(5.63), or the equivalent parameters $f_\epsilon$ and $f_p$ can
then be determined by using
$$
{\cal J}_l~~{\sf and~~any}~ |V_{ij}|;\eqno(5.67)
$$
i.e., the absolute value of any element of the experimentally
measured neutrino mapping matrix. All other remaining elements of
the neutrino mapping matrix may serve as part of the test of the
timeon model. Further discussions are given in Appendix D.

\section*{\Large \sf Acknowledgement}

We wish to thank N. Samios for discussions.

\vspace{1cm}

{\normalsize

\begin{center}
Table 1
\end{center}
$$
V_{l-{\sf map}}=\left[
\begin{array}{ccc}
\sqrt{\frac{2}{3}}\sin\frac{\alpha}{2}+\sqrt{\frac{1}{6}}\cos\frac{\alpha}{2}e^{i\beta}+
\sqrt{\frac{1}{6}}c_e
&\sqrt{\frac{1}{3}}(\sin\frac{\alpha}{2}-\cos\frac{\alpha}{2}e^{i\beta}-c_e)
& \sqrt{\frac{1}{2}}(-\cos\frac{\alpha}{2}e^{i\beta}+c_e)\\
\sqrt{\frac{2}{3}}\cos\frac{\alpha}{2}e^{-i\beta}-\sqrt{\frac{1}{6}}\sin\frac{\alpha}{2}+
\sqrt{\frac{1}{6}}c_\mu
&\sqrt{\frac{1}{3}}(\cos\frac{\alpha}{2}e^{-i\beta}+\sin\frac{\alpha}{2}-c_\mu)
&\sqrt{\frac{1}{2}}(\sin\frac{\alpha}{2}+c_\mu)~~\\
\frac{1}{m}(\sqrt{\frac{2}{3}}\chi_\epsilon-\sqrt{\frac{1}{6}}\chi_p)+\sqrt{\frac{1}{6}}
&\frac{1}{m}\sqrt{\frac{1}{3}}(\chi_\epsilon+\chi_p)-\sqrt{\frac{1}{3}}
&\sqrt{\frac{1}{2}}(\frac{1}{m}\chi_p+1)
\end{array}\right]
$$
}

\vspace{0.5cm}

Table 1. Lepton Mapping Matrix (neglecting $O(m^{-2})$), with the
parameters $\alpha=\alpha_l$ and $\beta=\beta_l$ given by (5.47);
$c_e,~c_\mu$ by (5.52), and $\chi_\epsilon,~\chi_p$ by (5.53) and
(2.36)-(2.37). In the limit $m\rightarrow \infty$,
$c_e,~c_\mu,~m^{-1}\chi_\epsilon$ and $m^{-1}\chi_p$ all become
$0$.

\newpage

\section*{\Large \sf Appendix A~~ Two Forms of Mass Matrix}

\noindent{\bf A.1 General Formulation}\\

As in (2.45), the mass matrix ${\cal M}$ and its related
Hamiltonian ${\cal H}$ of a Dirac field operator $\Psi$ with
$n$-generation components can be written as
$$
{\cal H}=\Psi^\dag {\cal M} \gamma_4 \Psi\eqno(A.1)
$$
in which
$$
{\cal M}={\cal M}^\dag,\eqno(A.2)
$$
denoting a hermitian matrix. Decompose $\Psi$ into a sum of
left-handed and right-handed parts:
$$
\Psi={\cal L} + {\cal R}\eqno(A.3)
$$
with
$$
{\cal L}=\frac{1}{2}(1+\gamma_5)\Psi~~~{\sf and}~~~{\cal
R}=\frac{1}{2}(1-\gamma_5)\Psi.\eqno(A.4)
$$
Correspondingly, (A.1) becomes
$$
{\cal H}={\cal L}^\dag{\cal M}\gamma_4{\cal R} +{\cal R}^\dag{\cal
M}\gamma_4{\cal L}.\eqno(A.5)
$$
Assume $n\geq 3$ and ${\cal M}$ to have an imaginary part so that
${\cal H}$ is $T$, $C$ and $CP$ violating.

A different form of an $n$-generation $T$ and $CP$ violating mass
Hamiltonian can be written in the form similar to (1.3), also with
a Dirac operator $\psi$ of $n$ components:
$$
H=\psi^\dag({\cal G}\gamma_4+i{\cal
F}\gamma_4\gamma_5)\psi,\eqno(A.6)
$$
where ${\cal G}$ and ${\cal F}$ are both $n$-dimensional hermitian
matrices,
$$
{\cal G}={\cal G}^\dag~~~{\sf and}~~~{\cal F}={\cal
F}^\dag.\eqno(A.7)
$$
For $n\geq 3$, ${\cal G}$ and ${\cal F}$ both nonzero, the
Hamiltonian $H$ is $T$, $P$ and $CP$ violating. As in (A.3)-(A.4),
we resolve $\psi$ in a similar form:
$$
\psi=L+R\eqno(A.8)
$$
with
$$
L=\frac{1}{2}(1+\gamma_5)\psi~~~{\sf
and}~~~R=\frac{1}{2}(1-\gamma_5)\psi.\eqno(A.9)
$$
Thus, (A.6) becomes
$$
H=L^\dag({\cal G}-i{\cal F})\gamma_4 R+R^\dag({\cal G}+i{\cal
F})\gamma_4L,\eqno(A.10)
$$
different from (A.5).

In the standard model, excluding the mass Hamiltonian, only the
left hand components of $\uparrow$ and $\downarrow$ quarks are
linked by their $W$-interaction. Hence, the right-hand component
${\cal R}$ or $R$ can undergo an independent arbitrary unitary
transformation. Because of this freedom, we can bring (A.10) into
the form (A.5), or vice versa, as is well known. We shall review
this equivalence, and then discuss how this equivalence can be
altered by imposing new restrictions on these matrices.

To show this, we begin with the form (A.10). Define
$$
M={\cal G}-i{\cal F}\eqno(A.11)
$$
and assume it to be nonsingular (i.e., the eigenvalues of $M^\dag
M$ are all nonzero.) On account of (A.7), the hermitian conjugate
of $M$ is
$$
M^\dag={\cal G}+i{\cal F}.\eqno(A.12)
$$
Since $MM^\dag$ is hermitian, there exists a unitary matrix $V_L$
that can diagonalize $MM^\dag$, with
$$
V_L^\dag MM^\dag V_L=m_D^2=~{\sf Diagonal}.\eqno(A.13)
$$
For every eigenvector $\phi$ of $MM^\dag$ with eigenvalue
$\lambda$, the corresponding vector $M^\dag \phi$ is an
eigenvector of $M^\dag M$ with the same eigenvalue $\lambda$.
Thus, $M^\dag M$ can also be diagonalized by another unitary
matrix $V_R$ as
$$
V_R^\dag M^\dag MV_R=m_D^2,\eqno(A.14)
$$
with $m_D^2$ the same diagonal matrix of (A.13).

Multiply (A.13) on the right by $m_D^{-1}$, it follows that
$$
V_L^\dag MV_R=m_D,\eqno(A.15)
$$
provided that we define
$$
V_R=M^\dag V_Lm_D^{-1}.\eqno(A.16)
$$
One can readily see that $V_L$ and $V_R$ thus defined satisfies
$V_L^\dag V_L=1$, $V_R^\dag V_R=1$ as well as (A.13) and (A.14).
Since $R$ can be transformed independently from $L$, we can
transform the $\psi$ field by
$$
L\rightarrow V_LL\eqno(A.17)
$$
and
$$
R\rightarrow V_RR.\eqno(A.18)
$$

Next let us examine the mass matrix ${\cal M}$ of (A.1)-(A.2).
Because ${\cal M}$ is hermitian, it can be diagonalized by a
single unitary transformation $V$, with the left-handed and
right-handed components of the field operator $\Psi$ undergoing
the \underline{same} transformation; i.e., in contrast to
(A.17)-(A.18), we have
$$
{\cal L}\rightarrow V{\cal L},\eqno(A.19)
$$
$$
{\cal R}\rightarrow V{\cal R}\eqno(A.20)
$$
and correspondingly
$$
{\cal H}\rightarrow \Psi^\dag m_D\gamma_4\Psi\eqno(A.21)
$$
with $m_D$ being the corresponding diagonal matrix. So far as the
mass matrices are concerned, we regard these two mass Hamiltonians
${\cal H}$ and $H$ as equivalent, if the diagonal matrix $m_D$ of
(A.21) has the same set of eigenvalues as those in (A.15). In this
case, we can without loss of generality set
$$
{\cal M}^2=({\cal G}-i{\cal F})({\cal G}+i{\cal F})\eqno(A.22)
$$
and
$$
V=V_L;\eqno(A.23)
$$
hence (A.13) becomes
$$
V^\dag{\cal M}^2V=m_D^2\eqno(A.24)
$$
and therefore
$$
V^\dag{\cal M} V=m_D.\eqno(A.25)
$$
(Note that ${\cal M}\neq M$ or $M^\dag$, even though ${\cal M}^2=
MM^\dag$.)\\

\newpage

\noindent{\bf A.2 Restricted Class with ${\cal G}$ and
${\cal F}$ Real}\\

We will now discuss theories in which both matrices ${\cal G}$ and
${\cal F}$ are real; i.e.,
$$
{\cal G}={\cal G}^*~~{\sf and}~~{\cal F}={\cal F}^*.\eqno(A.26)
$$
Since ${\cal G}$ and ${\cal F}$ are also hermitian; they must both
be symmetric matrices. In $n$-dimension, each of these matrices
can carry $\frac{1}{2}n(n+1)$ independent real parameters, giving
a total of $n(n+1)$ real parameters. On the other hand, ${\cal M}$
being a single hermitian matrix consists of only $n^2$ real
parameters. Thus, knowing ${\cal G}$ and ${\cal F}$, by using
(A.22), we can always determine uniquely the corresponding ${\cal
M}$, but not the converse, by expanding in power series as
follows.

Decompose the hermitian ${\cal M}$ into its real and imaginary
parts:
$$
{\cal M}=R+iI.\eqno(A.27)
$$
with $R$ and $I$ both real; hence, $R$ is symmetric and $I$
antisymmetric. On account of (A.26), the real part of (A.22) is
$$
R^2-I^2={\cal G}^2+{\cal F}^2,\eqno(A.28)
$$
and the imaginary part is
$$
\{R,~I\}=[{\cal G},~{\cal F}].\eqno(A.29)
$$
In what follows, we assume that ${\cal G}$ and ${\cal F}$ are both
known, as in the case when the mass matrix is given by (1.3). In
addition, ${\cal F}$ can be regarded as small compared to ${\cal
G}$. Hence, we can expend $R$ and $I$ in powers of ${\cal F}$.
Write
$$
R={\cal G}+R_2+R_4+R_6+\cdots\eqno(A.30)
$$
and
$$
I=I_1+I_3+I_5+\cdots,\eqno(A.31)
$$
with $R_n$ and $I_m$ to be of the order of ${\cal F}^n$ and ${\cal
F}^m$ respectively. Eqs.(A.28) and (A.29) give
$$
\{{\cal G},~I_1\}=[{\cal G},~{\cal F}],~~~~~~
$$
$$
\{{\cal G},~R_2\}={\cal F}^2+I_1^2,~~~~~
$$
$$
\{{\cal G},~I_3\}=-\{R_2,~I_1\},\eqno(A.32)
$$
$$
\{{\cal G},~R_4\}=\{I_1,~I_3\}-R_2^2,~~{\sf etc.}
$$
As noted before, so far as these mass matrices are concerned, the
two formalisms (A.1) and (A.6) are regarded as equivalent to each
other, provided that (A.28) and (A.29) hold. Then (A.32) gives the
conditions determining the series expansions (A.30)-(A.31) of $R$
and $I$ in terms of ${\cal G}$ and ${\cal F}$.

It is convenient to write ${\cal G}$ in terms of its eigenvalues
$\nu,~\mu,~m$ and their corresponding eigenvectors $\epsilon,~p,~P$
(as in (2.14)):
$$
{\cal G}=\nu\epsilon\tilde{\epsilon}+\mu p\tilde{p} +m
P\tilde{P}.\eqno(A.33)
$$
Let $\vec{{\cal A}}$ be a vector whose $k^{th}$ component is given
by the $(i,~j)^{th}$ component of the commutator between ${\cal G}$
and ${\cal F}$:
$$
[{\cal G},~{\cal F}]_{ij}=\epsilon_{ijk}{\cal A}_k\eqno(A.34)
$$
with $\epsilon _{ijk}=\pm 1$ depending on $(ijk)$ being an even or
odd permutation of $(1,2,3)$, and $0$ otherwise. From (A.33) and
(2.18)-(2.19), we can readily verify that
$$
{\cal A}_k=\tau\bigg(\nu(\hat{\epsilon}\cdot
\hat{f})(\hat{\epsilon}\times\hat{f})+ \mu(\hat{p}\cdot
\hat{f})(\hat{p}\times\hat{f}) +m(\hat{P}\cdot
\hat{f})(\hat{P}\times\hat{f})\bigg)_k.\eqno(A.35)
$$

Since $I$ is antisymmetric, so is $I_1$. Write  its $(ij)$th
component as
$$
(I_1)_{ij}=\epsilon_{ijk}J_k.\eqno(A.36)
$$
By using the first equation of (A.32) with (A.33)-(A.36), it can be
readily verified that $\vec{J}$ is related to $\vec{A}$ by
$$
\vec{J}\cdot\hat{\epsilon}=(\mu+m)^{-1}\vec{{\cal A}}\cdot
\hat{\epsilon}
$$
$$
\vec{J}\cdot\hat{p}=(m+\nu)^{-1}\vec{{\cal A}}\cdot
\hat{p}\eqno(A.37)
$$
and
$$
\vec{J}\cdot\hat{P}=(\nu+\mu)^{-1}\vec{{\cal A}}\cdot \hat{P}.
$$
In the same way, we can solve for $R_2$, $I_3$, $\cdots$.

\newpage

\section*{\Large \sf Appendix B~~ Proof of Eq.(3.34) for Jarlskog Invariant}

\noindent{\bf B.1 Definitions, Corollaries and Conventions}\\

Let $V=(V_{i\alpha})$ be a $3\times 3$ real orthogonal matrix with
positive determinant, and indices $i$ and $\alpha=1,~2$ and $3$;
hence,
$$
V=V^*,~~V^{-1}=\tilde{V}~~{\sf and}~~|V|=1.\eqno(B.1)
$$
Given any value of $i$, define $i',~i''$ by
$$
\epsilon_{ii'i''}=1,\eqno(B.2)
$$
so that $(i,~i',~i'')$ is a cyclic permutation of $(1,~2,~3)$. We
shall use the same definition for each of such similar indices
$j,~k,~l,~\alpha,~\beta,~\gamma$. Thus, for a given $j$, the
corresponding $j'$ and $j''$ satisfy
$$
\epsilon_{jj'j''}=1,\eqno(B.3)
$$
and likewise
$$
\epsilon_{kk'k''}=\epsilon_{ll'l''}=\epsilon_{\alpha\alpha'\alpha''}
=\epsilon_{\beta\beta'\beta''}=\epsilon_{\gamma\gamma'\gamma''}=1.
$$
Furthermore, for any pair $i,~\alpha$, we have by the expression
for $V^{-1}$
$$
\left|
\begin{array}{cc}
V_{i'\alpha'} & V_{i'\alpha''}\\
V_{i''\alpha'} & V_{i''\alpha''}
\end{array}
\right| =|V|(V^{-1})_{\alpha i}=V_{i\alpha}\eqno(B.4)
$$
on account of (B.1). This identity will enable us to reduce
certain quartic products of $V_{i\alpha}$ to triple products of
$V_{i\alpha}$, as we shall see.\\

\noindent{\bf B.2 Jarlskog Invariant}\\

Similar to the relation between $V$ of (3.32) and
$$
U=U_{CKM}\eqno(B.5)
$$
of (3.33), we define $W=(W_{i\alpha})$ through
$$
U=V+i\tau_qW\eqno(B.6)
$$
where $W$ has the form
$$
W_{i\alpha}=\sum_{j \neq i}F_{ij}V_{j\alpha} -\sum_{\beta
\neq\alpha}{\cal F}_{\alpha\beta}V_{i\beta}.\eqno(B.7)
$$
For our purpose here, it is necessary to specify only that $F$ and
${\cal F}$ are real and symmetric; i.e.,
$$
F_{ij}=F_{ji}=F_{ij}^*\eqno(B.8)
$$
and
$$
{\cal F}_{\alpha\beta}={\cal F}_{\beta\alpha}={\cal
F}_{\alpha\beta}^*.\eqno(B.9)
$$
We also define for any particular pair of indices $k$ and
$\gamma$,
$$
J= U_{k\gamma} U_{k'\gamma'}U_{k\gamma'}^*
U_{k'\gamma}^*\eqno(B.10)
$$
and
$$
J_0= V_{k\gamma} V_{k'\gamma'}V_{k\gamma'}^*
V_{k'\gamma}^*.\eqno(B.11)
$$
Thus, substituting (B.6) into (B.10), we find, to first order in
$\tau_q$,
$$
J-J_0=i\tau_q\Delta_{k\gamma}\eqno(B.12)
$$
where
$$
\Delta_{k\gamma}=J_0\bigg(\frac{W_{k\gamma}}{V_{k\gamma}}+
\frac{W_{k'\gamma'}}{V_{k'\gamma'}}-\frac{W_{k\gamma'}}{V_{k\gamma'}}
-\frac{W_{k'\gamma}}{V_{k'\gamma}}\bigg).\eqno(B.13)
$$
Note that $k,~\gamma$  are subject to the cyclic convention
typified by (B.2). [It will turn out that $\Delta_{k\gamma}$ is
independent of the choice of $k$ and $\gamma$, even though this is
not assumed.]

By substituting (B.7) into (B.13), we must obtain an expression of
the form
$$
\Delta_{k\gamma}=\sum_l A_{l''}F_{ll'}+ \sum_\lambda {\cal
A}_{\lambda''}{\cal F}_{\lambda\lambda'}\eqno(B.14)
$$
where the $A's$ and ${\cal A}'s$ are to be determined. From (B.7),
(B.13) and (B.14), we see that each $A$ is made of terms having
the form $J_0V_{j\alpha}/V_{i\alpha}$. Consider $A_k$: in (B.14)
we must put $l''=k$; hence $l,~l'$ are $k',~k''$ in some order.
Thus in (B.7), $i$ is either $k'$ or $k''$. But the index $k''$
does not occur in (B.13). Therefore, we have $i=k', j=k''$ and
$\alpha=\gamma$ or $\gamma'$. Therefore,
$$
A_k=J_0\bigg(0+\frac{V_{k''\gamma'}}{V_{k'\gamma'}}-0-
\frac{V_{k''\gamma}}{V_{k'\gamma}}\bigg)
$$
$$
=V_{k\gamma}V_{k\gamma'}\left|
\begin{array}{cc}
V_{k'\gamma} & V_{k'\gamma'}\\
V_{k''\gamma} & V_{k''\gamma'}
\end{array}
\right| =V_{k\gamma}V_{k\gamma'}V_{k\gamma''}\eqno(B.15)
$$
on account of (B.4). Likewise,
$$
A_{k'}=V_{k'\gamma}V_{k'\gamma'}V_{k'\gamma''}.\eqno(B.16)
$$

For $k''$, the calculation is different, even though the result
will be similar. Note that in (B.14) $l,~l'$ can be $k,~k'$ in
either order, so that in (B.7) $i$ and $j$ can also be $k,~k'$ in
either order. Thus, we now have four terms instead of two:
$$
A_{k''}=J_0\bigg(\frac{V_{k'\gamma}}{V_{k\gamma}}+
\frac{V_{k\gamma'}}{V_{k'\gamma'}}
-\frac{V_{k'\gamma'}}{V_{k\gamma'}}-
\frac{V_{k\gamma}}{V_{k'\gamma}}\bigg)
$$
$$
=V_{k'\gamma}V_{k'\gamma'}\left|
\begin{array}{cc}
V_{k'\gamma} & V_{k'\gamma'}\\
V_{k\gamma} & V_{k\gamma'}
\end{array}
\right| + V_{k\gamma}V_{k\gamma'}\left|
\begin{array}{cc}
V_{k\gamma'} & V_{k\gamma}\\
V_{k'\gamma'} & V_{k'\gamma}
\end{array}
\right|
$$
$$
=-V_{k'\gamma}V_{k'\gamma'}V_{k''\gamma''}-V_{k\gamma}V_{k\gamma'}V_{k\gamma''}
=+V_{k''\gamma}V_{k''\gamma'}V_{k''\gamma''}\eqno(B.17)
$$
since $V$ is orthogonal.

Similar formulas are obtained for ${\cal A}_\gamma$, ${\cal
A}_\gamma'$, ${\cal A}_\gamma''$ with a change of sign. Since
(B.15)-(B.17) all have the same form, the choice of $k,~\gamma$ in
(B.10) and (B.11) is immaterial  and we have
$$
J=J_0+i\tau_q\Delta\eqno(B.18)
$$
where
$$
\Delta=\sum_l F_{ll'}\prod_\alpha V_{l''\alpha}-\sum_\lambda {\cal
F}_{\lambda\lambda'}\prod_i V_{i\lambda''}.\eqno(B.19)
$$

\noindent{\bf B.3 Applications to quarks}\\

By consulting (3.7)-(3.15) and (3.17)-(3.25), we find that
$$
F_{ij}=\frac{f_{i}f_{j}}{m_i+m_j},~~~ {\cal F}
_{\alpha\beta}=\frac{
\bar{f}_{\alpha}\bar{f}_\beta}{\mu_\alpha+\mu_\beta}\eqno(B.20)
$$
where
$$
f_1=(f_\epsilon)_\uparrow,~~f_2=(f_p)_\uparrow,~~f_3=(f_P)_\uparrow,
$$
$$
\bar{f}_1=(f_\epsilon)_\downarrow,~~
\bar{f}_2=(f_p)_\downarrow,~~\bar{f}_3=(f_P)_\downarrow\eqno(B.21)
$$
with
$$
m_1=m_u=0,~~m_2=m_c,~~m_3=m_t,
$$
$$
\mu_1=m_d=0,~~\mu_2=m_s,~~\mu_3=m_b.\eqno(B.22)
$$
Then (3.34) is seen to be (B.14) if we identify
$$
A_c=A_3,~~A_t=A_2,~~B_\uparrow=A_1,
$$
$$
A_s={\cal A}_3,~~A_b={\cal A}_2~~{\sf and}~~B_\downarrow={\cal
A}_1.\eqno(B.23)
$$
With these identifications, (3.38)-(3.40) are (B.15)-(B.17) and
(3.35)-(3.37) are the corresponding formulas for ${\cal A}_1,~{\cal
A}_2$ and ${\cal A}_3$.\\


\section*{\Large \sf Appendix C~~ Phase Convention in $V_{l-{\sf map}}$}

~~In order to compare the lepton mapping matrix $V_{l-{\sf map}}$
given in Table~1 and the experimentally measured neutrino mapping
matrix
$$
U_\nu=(U_{ij}),\eqno(C.1)
$$
there are certain phase conventions.  In definition of
$$
V_{l-{\sf map}}=\left(
\begin{array}{ccc}
<\nu_1|e>&<\nu_2|e>&<\nu_3|e>\\
<\nu_1|\mu>&<\nu_2|\mu>&<\nu_3|\mu>\\
<\nu_1|\tau>&<\nu_2|\tau>&<\nu_3|\tau>
\end{array}\right)\eqno(C.2)
$$
given by Table~1, we made arbitrary phase choices of these
statevectors. This allows us to consider the following
transformations:
$$
\left(
\begin{array}{c}
|e>\\
|\mu>\\
|\tau>
\end{array}\right)\rightarrow\Omega_l
\left(
\begin{array}{c}
|e>\\
|\mu>\\
|\tau>
\end{array}\right)\eqno(C.3)
$$
and
$$
\left(
\begin{array}{c}
|\nu_1>\\
|\nu_2>\\
|\nu_3>
\end{array}\right)\rightarrow\Omega_\nu
\left(
\begin{array}{c}
|\nu_1>\\
|\nu_2>\\
|\nu_3>
\end{array}\right),\eqno(C.4)
$$
where
$$
\Omega_l=\left(
\begin{array}{ccc}
e^{i\xi_e}&0&0\\
0&e^{i\xi_\mu}&0\\
0&0&e^{i\xi_\tau}
\end{array}\right)\eqno(C.5)
$$
and
$$
\Omega_\nu=\left(
\begin{array}{ccc}
e^{i\eta_1}&0&0\\
0&e^{i\eta_2}&0\\
0&0&e^{i\eta_3}
\end{array}\right).\eqno(C.6)
$$
The experimentally measured neutrino mapping matrix $U_\nu$ is
related to $V_{l-{\sf map}}$ by
$$
U_\nu=\Omega_l V_{l-{\sf map}}\Omega_\nu^{-1},\eqno(C.7)
$$
with the conditions
$$
|U_\nu|=1\eqno(C.8)
$$
and the following four matrix elements of $U_\nu$,
$$
U_{11},~U_{12},~U_{23}~~{\sf and }~~U_{33}\eqno(C.9)
$$
all real and positive. Thus, (C.8) and (C.9) determine five of the
six phase factors in (C.5)-(C.6). The remaining one
$$
\xi_e+\xi_\mu+\xi_\tau+\eta_1+\eta_2+\eta_3\eqno(C.10)
$$
does not appear in $U_\nu$.\\


\section*{\Large \sf Appendix D~~ Test for Leptonic System}

~~Following the discussion given in Remarks at the end of
Section~5, we may consider
$$
{\cal J}_l~~{\sf and~~}|V_{11}|\eqno(D.1)
$$
as an example of (5.67), and two other members, say
$$
|V_{12}|~~{\sf and}~~|V_{23}|.\eqno(D.2)
$$
From Table~1 and noting that $c_e$ and $c_\mu$ are $O(m^{-1})$, we
have
$$
|V_{11}|^2=
\frac{2}{3}\sin^2\frac{\alpha}{2}+\frac{1}{6}\cos^2\frac{\alpha}{2}
+\frac{2}{3}\sin\frac{\alpha}{2}\cos\frac{\alpha}{2}\cos\beta
$$
$$
~~~~+\frac{1}{3}Re\bigg[c_e^*(2\sin\frac{\alpha}{2}
+e^{i\beta}\cos\frac{\alpha}{2})\bigg],
$$
$$
|V_{12}|^2=
\frac{1}{3}\sin^2\frac{\alpha}{2}+\frac{1}{3}\cos^2\frac{\alpha}{2}
-\frac{2}{3}\sin\frac{\alpha}{2}\cos\frac{\alpha}{2}\cos\beta
$$
$$
~~~~-\frac{2}{3}Re\bigg[c_e^*(\sin\frac{\alpha}{2}
-e^{i\beta}\cos\frac{\alpha}{2})\bigg]\eqno(D3)
$$
and
$$
|V_{23}|^2=
\frac{1}{2}\sin^2\frac{\alpha}{2}+Re[c_\mu^*\sin\frac{\alpha}{2}],
~~~~~~~~~~~~~~~~~~~~
$$
in which $\alpha=\alpha_l$, $\beta=\beta_l$ and all $O(m^{-2})$
terms are not included. Using (5.52)-(5.53) and noting that, to
our approximation, (2.36)-(2.37) yield
$$
\chi=\left(
\begin{array}{c}
\chi_\epsilon\\
\chi_p
\end{array}\right)=-i\tau f_P
\left(
\begin{array}{c}
f_\epsilon\\
f_p
\end{array}\right);\eqno(D.4)
$$
therefore, in accordance with (5.52)-(5.53)
$$
\begin{array}{ll}
&c_e^*=i\frac{\tau}{m}f_P(f_\epsilon\sin\frac{\alpha}{2}
-f_pe^{-i\beta}\cos\frac{\alpha}{2})~~~~~~~~~~~~~~~~~~~~~~~~~~~~~~~~~\\
{\sf and}~~~~~~~~~~~~~~~~~~~~~&~~~~~~~~~~~~~~~~~~~~~~~~~~~~~~~~~~
~~~~~~~~~~~~~~~~~~~~~~~~~~~(D.5)\\
&c_\mu^*=i\frac{\tau}{m}f_P(f_\epsilon
e^{i\beta}\cos\frac{\alpha}{2}
+f_p\sin\frac{\alpha}{2})~~~~~~~~~~~~~~~~~~~~~~~~~~~~~~~~~\\
\end{array}
$$
where
$$
\tau=\tau_l.\eqno(D.6)
$$
From (D.3) and (D.5) and eliminating $f_P$ by (2.21), we find
$$
|V_{11}|^2=
\frac{5}{12}-\frac{1}{4}\cos\alpha+\frac{1}{3}\sin\alpha\cos\beta~~~~~~~~~~~~~~
$$
$$
~~~~~~~~-\frac{1}{6}\frac{\tau}{m}(1-f_\epsilon^2-f_p^2)^{\frac{1}{2}}
(f_\epsilon+2f_p)\sin\alpha\sin\beta,
$$
$$
|V_{12}|^2=
\frac{1}{3}-\frac{1}{3}\sin\alpha\cos\beta~~~~~~~~~~~~~~~~~~~~~~~~~~~~~~
$$
$$
~~~~~~-\frac{1}{3}\frac{\tau}{m}(1-f_\epsilon^2-f_p^2)^{\frac{1}{2}}
(f_\epsilon-f_p)\sin\alpha\sin\beta\eqno(D.7)
$$
and
$$
|V_{23}|^2=
\frac{1}{4}-\frac{1}{4}\cos\alpha~~~~~~~~~~~~~~~~~~~~~~~~~~~~~~~~~~~~
$$
$$
-\frac{1}{2}\frac{\tau}{m}(1-f_\epsilon^2-f_p^2)^{\frac{1}{2}}
f_\epsilon \sin\alpha\sin\beta,~~
$$
in which $\alpha=\alpha_l$ and $\beta=\beta_l$ are defined by
(5.47). Thus, $|V_{11}|^2$, $|V_{12}|^2$ and $|V_{23}|^2$ are all
functions of $m,~\mu,~\tau,~f_\epsilon$ and $f_p$.

Finally, the Jarlskog invariant is given by (5.58)-(5.59). These
expressions provide the explicit forms for the four observables in
(D.1)-(D.2). Together with (5.45) and (5.64) for the masses of $e$,
$\mu$ and $\tau$, we have seven observables in terms of five
parameters.\\


\vspace{2cm}

\centerline{\epsfig{file=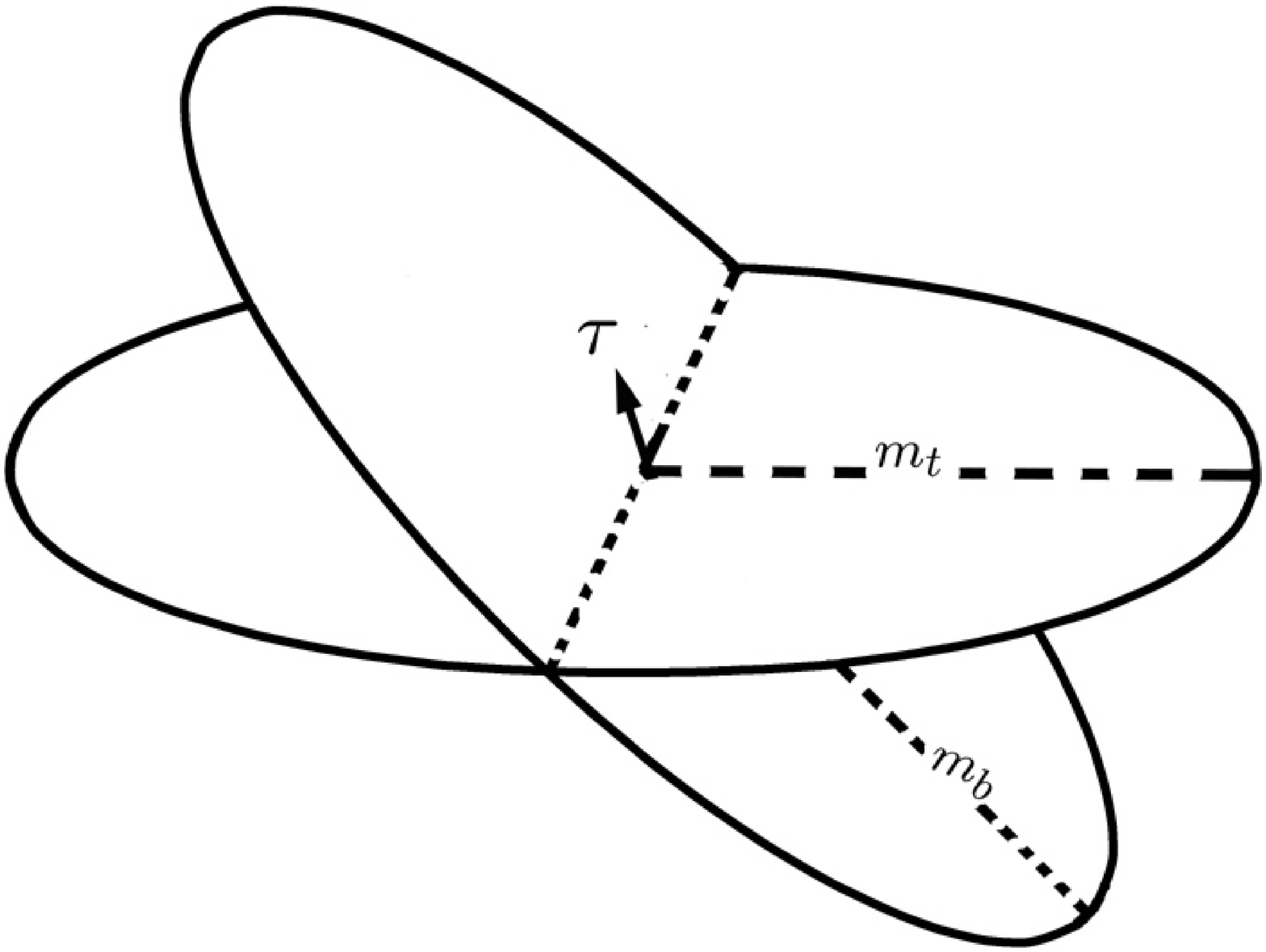,width=12cm}}

\vspace{.5cm}

\noindent Figure~1. A schematic drawing of the quark mass matrix
$M_{\uparrow/\downarrow}=G_{\uparrow/\downarrow}\gamma_4 +
iF\gamma_4 \gamma_5$. The vibration of $\tau(x)$ is timeon\\

\newpage

\centerline{\epsfig{file=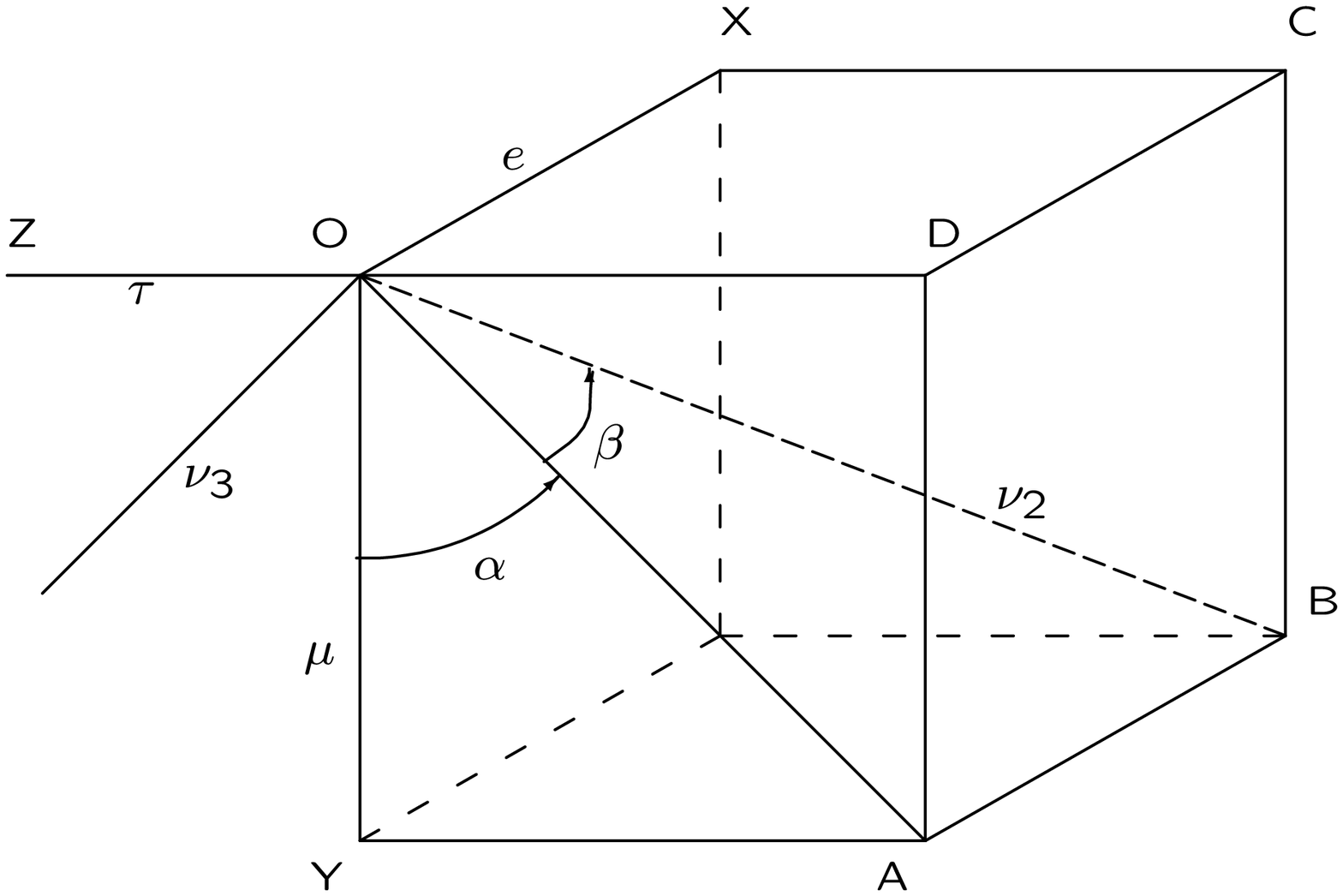,width=13cm}}

\vspace{.5cm}

\noindent Figure~2. Geometric Representation of Harrison, Perkins,
Scott Transformation (5.1) and (5.6)-(5.8)\\

\noindent The axes $\overline{OX},~\overline{OY}$ and
$\overline{OZ}= -\overline{OD}$ represent $e,~\mu$ and $\tau$. An
$\alpha=45^0$ left-hand rotation $(V_\downarrow)_0$ along $e$ takes
$\tau$ to $\nu_3$ and $\mu$ to $\overline{OA}$. A second
$\beta=\sin^{-1}\sqrt{\frac{1}{3}}$ left-hand rotation
$(V_\uparrow)_0^\dag$ along $\nu_3$ takes $e$ to $\nu_1$ and
$\overline{OA}$ to $\nu_2$ which is along $\overline{OB}$. The
hidden bases are $l_1=e$, $l_2\parallel\overline{OA}$ and
$l_3=\nu_3$.

\newpage

\section*{\Large \sf References}

\noindent [1] R. Friedberg and T. D. Lee, Ann. Phys. {\bf 323}(2008)1677\\

\noindent [2] T. D. Lee, Phys. Reports {\bf 9}(1974)143\\

\noindent [3] C. Jarlskog, Phys. Rev. {\bf D35}(1987)1685\\

\noindent [4] M. Gell-Mann and M. Levy, Nuovo Cimento 16(1960)705

~N. Cabibbo, Phys. Rev. Lett., 10(1963)531\\

\noindent [5] M. Kabayashi and T. Maskawa, Prog. Th. Phys.
49(1973)652\\

\noindent [6] We wish to thank C. Q. Geng for calling our attention
to the high mass of

~~~~the timeon quantum because of its
flavor-changing properties.\\

\noindent [7] S. Eidelman et.al. Particle Data Group, Phys. Lett. {\bf B592}(2004)1\\

\noindent [8] P. F. Harrison, D. H. Perkins and W. G. Scott,

~~~~~~~~~~Phys. Lett. B530, 167(2002).

~Z. Z. Xing, Phys. Lett. B533, 85(2002);

~P. F. Harrison and W. G. Scott, Phys. Lett. B535, 163(2002);

~X. G. He and A. Zee, Phys. Lett. B560, 87(2003).\\

\noindent [9] G. L. Fogli, et al., Phys. Rev. D78, 033010(2008);

~G. L. Fogli, et al., Phys. Rev. Lett. 101, 141801(2008).\\

\noindent [10] Z. Z. Xing, 34th International Conference on High
Energy Physics,

~~~~Philadelphia(2008), arXiv:0810.1421v2[hep-ph]2 Jan.(2009).

\end{document}